\documentclass[aps,prb,twocolumn,groupedaddress]{revtex4}
\usepackage{graphicx,amssymb,bm}

\begin{document}
\title{Unpinning triggers for superfluid vortex avalanches}
\author{L. Warszawski}
\email[]{lila@unimelb.edu.au}
\author{A. Melatos}
\affiliation{School of Physics, The University of Melbourne, Parkville 3010, Victoria, Australia}
\author{N.~G. Berloff}
\affiliation{Department of Applied Mathematics and Theoretical Physics, University of Cambridge, Cambridge, CB3 0WA, United Kingdom}
\date{\today}
\begin{abstract}
The pinning and collective unpinning of superfluid vortices in a decelerating container is a key element of the canonical model of neutron star glitches and laboratory spin-down experiments with helium II.  Here the dynamics of vortex (un)pinning is explored using numerical Gross-Pitaevskii calculations, with a view to understanding the triggers for catastrophic unpinning events (vortex avalanches) that lead to rotational glitches.  We explicitly identify three triggers:  rotational shear between the bulk condensate and the pinned vortices, a vortex proximity effect driven by the repulsive vortex-vortex interaction, and sound waves emitted by moving and repinning vortices.  So long as dissipation is low, sound waves emitted by a repinning vortex are found to be sufficiently strong to unpin a nearby vortex.  For both ballistic and forced vortex motion, the maximum inter-vortex separation required to unpin scales inversely with pinning strength.
\end{abstract}

\maketitle
\section{\label{sec:intro}Introduction}
Superfluid vortices carrying quanta of circulation are pinned by lattice-scale impurities \citep{Jones:1997p26,Jones:1998p34,Donati:2003p97,Donati:2006p32,Avogadro:2007p51,Avogadro:2008p29} and/or macroscopic container defects \citep{DeBlasio:1998p127,Jones:2003p10}, because it is energetically favourable to superpose the empty vortex core and the condensate-excluding impurity.  In a neutron star, vortex pinning prevents the superfluid core from imitating the smooth electromagnetic deceleration of the crust (its container). The rotational shear that accumulates thus is corrected in discrete, randomly timed events, known as glitches \citep{Anderson:1975p84,Alpar:1981p18,Dodson02,Melatos:2008p204}.  Similar spasmodic spin-up events have also been observed in laboratory experiments with helium II in vessels of varying geometry and constitution \cite{Tsakadze:1975p5864,Andronikashvili:1979p4718}.

The statistics of neutron star glitches suggest that the underlying physics is a collective, self-organising process \citep{Melatos:2008p204}, akin to grain avalanches in sand pile experiments  \citep{Wiesenfeld:1989p44,Field:1995p155,Morley:1996p2128} and flux-tube avalanches in type II superconductors \citep{Olson:1996p3506,Bassler:1998p8} exposed to a variable applied magnetic field.  A small number of unpinnings is not sufficient, however, to cause a glitch in the above systems.  For example, in a neutron star glitch, anywhere from $10^{7}$ to $10^{14}$ vortices (out of $\sim 10^{18}$ in total) unpin simultaneously.  As the events are driven by global shear and threshold triggered in the canonical model, it is surprising that they are neither periodic in time nor equal in magnitude.  In fact, the probability distribution function (pdf) of glitch sizes has been shown to follow a power law, and the waiting-time pdf is well represented by an exponential \citep{Melatos:2008p204,Warszawski:2008p4510}.  Such statistical distributions are characteristic of a self-organized critical system \citep{Wiesenfeld:1989p44}.  In addition to unpinning collectively, in order to explain the observed glitch sizes the vortices must pass over many ($\sim 10^6$) nuclear lattice pinning sites as they move radially out before repinning.  This behaviour is not understood.

In order to catalyse an avalanche of simultaneous unpinnings, leading to abrupt acceleration of the container, the unpinning of a single vortex must raise the probability of other vortices (near or distant neighbours) unpinning.  In this paper, we identify two such \emph{knock-on} mechanisms, which operate in conjunction with the stochastic, non-cooperative unpinning driven by the global shear:  (1) radiation of sound waves, when a vortex moves and repins; and (2) a vortex proximity effect, when unpinned vortices approach adjacent pinned vortices.

Certain local aspects of the unpinning dynamics have been studied by other authors, but not the collective phenomenon of vortex avalanches.  The transition of a vortex lattice from an Abrikosov to a pinning-dominated configuration was studied in detail by \cite{Sato:2007p8103} and \cite{Yasunaga:2007p8108}, using the lattice energy as a diagnostic of the extent to which the vortices are pinned.  These authors did not discuss unpinning under global shear or track individual vortex dynamics.  Experiments reported in \textcite{Hakonen:1998p10787,Varoquaux:1998p2402} and \textcite{Varoquaux200087} quantify the velocity shear without contemplating knock on.  In the neutron star context, \textcite{Link:2009p9063} calculated the critical unpinning flow using the vortex-line equation of motion (Schwarz's equation) with and without dissipation in a random pinning potential.  Vortex-vortex interactions are built in via the Magnus force; acoustic radiation is not considered.  The spiral trajectory of a repinning vortex was also described hydrodynamically by \textcite{Sedrakian:1995MNRAS}.

In this paper we perform a set of numerical experiments, based on solutions of the time-dependent Gross-Pitaevskii equation (GPE), aimed at demonstrating the ability of sound waves, vortex-vortex proximity, and global velocity shear to unpin vortices.  The outcomes of these investigations will be employed in a future paper to inform the microscopic rules of an automaton model of neutron star glitches.   Section~\ref{GPE} briefly describes the numerics.  In Sec.~\ref{sec:shear} we measure the persistence of pinning in a velocity shear growing at a constant rate.  We study the energetics of the unpinning process and find evidence of acoustic radiation from the unpinned vortex.  Maintaining the same geometry, we then study the unpinning capacity of sound pulses in Sec.~\ref{sec:acoustic}.  First we artificially generate a pulse (Sec.~\ref{subsec:art_sound}), which unpins a nearby vortex.  We then measure the acoustic radiation from a repinning vortex (Sec.~\ref{subsec:vortex_sound}), which is then harnessed to unpin another nearby vortex (Sec.~\ref{subsec:sound_knock}).  Section~\ref{sec:prox} documents a series of experiments designed to assess the role of vortex proximity in unpinning.  Proximity between vortices is first arranged by forcibly dragging one vortex towards another, pinned vortex (Sec.~\ref{subsec:proxdrag}).  We then describe the unpinning of a vortex resulting from the ballistic (\emph{i.e.} free) approach of another, recently unpinned vortex (Sec.~\ref{subsec:ballistic}).  We relate our findings in Sec.~\ref{sec:shear}--\ref{sec:prox} to possible triggers for vortex avalanches in neutron star glitches in the concluding section (Sec.~\ref{sec:concindiv}).

\section{\label{GPE}Experimental setup}
\begin{figure*}
\begin{center}
\includegraphics[scale=.575,angle=90]{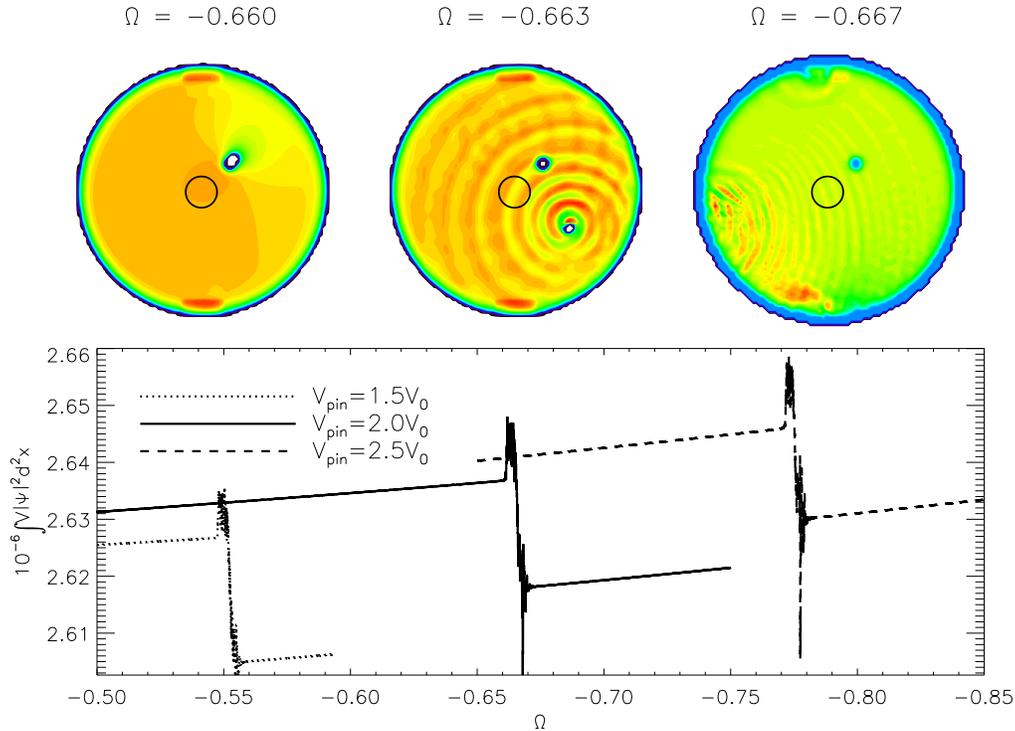}
\end{center}
\caption{\emph{Top}:  Density snapshots tracking the unpinning of a vortex due to global differential rotation, with $\Omega=\Omega_0+\dot{\Omega}t$, $\Omega_0=0.0$, and $\dot{\Omega}=-10^{-3}$.  The pinning site is at (1.7,1.7) and $R=8.5$.  The colour plots of condensate density $|\psi|^2$ are snapshots taken at $\Omega=-0.660$ (top left), $-0.663$ (top centre) and $-0.667$ (top right) for a pinning site of strength $V_i=2V_0$ ($V_0 = 60$).  The range of densities represented in the colour plots is $0.8|\psi|^2_{\rm{max}}$ to $|\psi|^2_{\rm{max}}$, so as to make low-amplitude sound waves visible.  The \emph{black circles} at the centre of each snapshot indicate the region in which kinetic energy is calculated in Fig.~\ref{fig:diff_kin}.  \emph{Bottom}:  Total potential energy $\int V|\psi|^2d^2x$ versus $\Omega$ for pinning sites with strength $V_i=1.5V_0,~2.0V_0,~\rm{and}~2.5V_0$ (dotted, solid and dashed curves) respectively.}
\label{fig:diff_latt}
\end{figure*}

\begin{figure*}
\begin{center}
\includegraphics[scale=.575,angle=90]{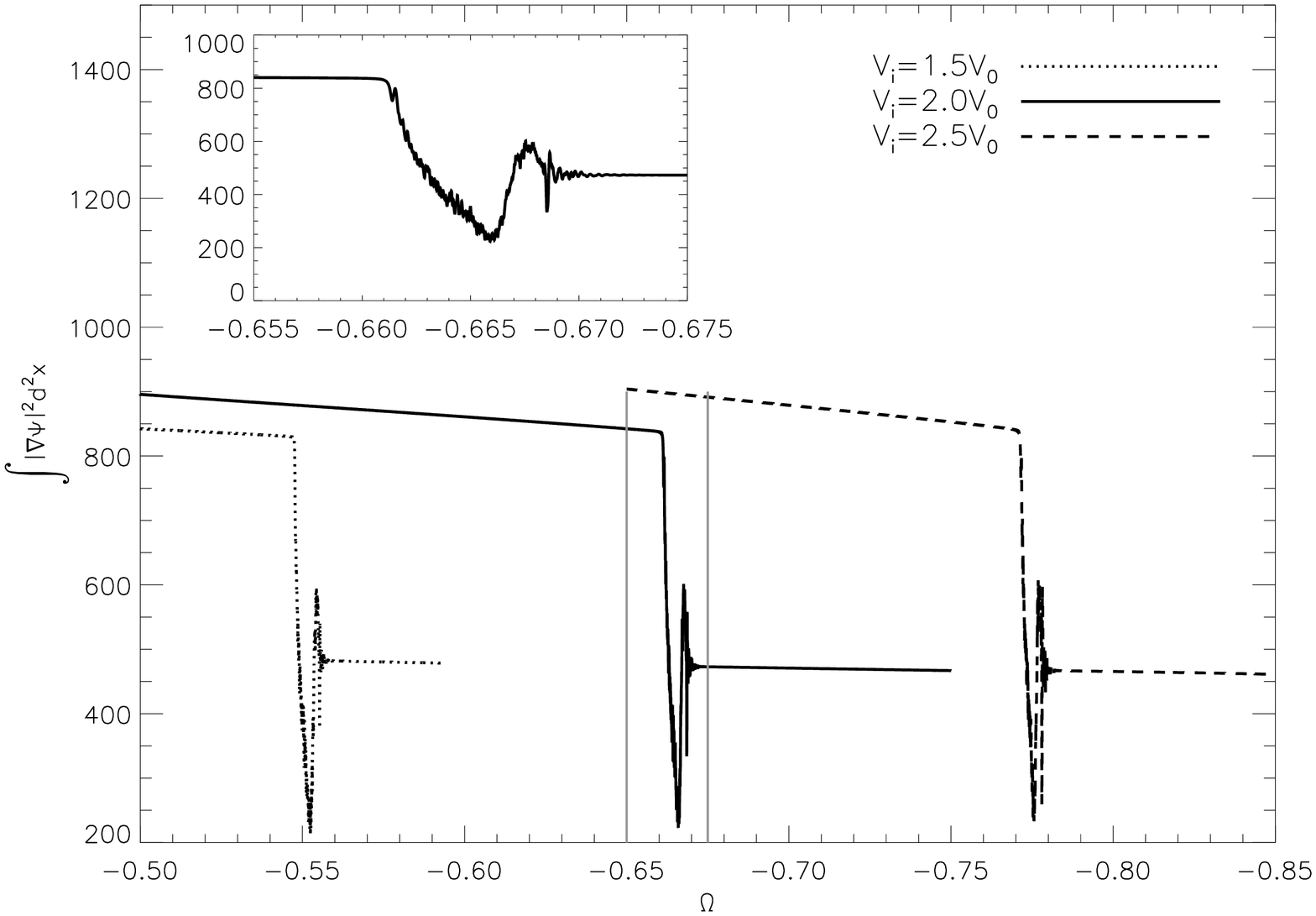}
\end{center}
\caption{Kinetic energy $\int |\nabla\psi|^2d^2x$ in a unit disk centred at (0,0) (depicted in the top panels of Fig.~\ref{fig:diff_latt}), for pinning sites with strength $V_i=1.5V_0,~2.0V_0,~\rm{and}~2.5V_0$ (dotted, solid and dashed curves respectively) for the experiment described by Fig.~\ref{fig:diff_latt}.  The inset zooms in on the region $-0.675\leq \Omega\leq -0.65$ (indicated by the vertical grey lines) for $V_i=2.0V_0$.}
\label{fig:diff_kin}
\end{figure*}

We begin by describing the numerical framework.  We employ the time-dependent, two-dimensional GPE to model a zero-temperature condensate inside a rotating, circular container.  The condensate order parameter $\psi(\mathbf{x},t)$, in a potential $V(\mathbf{x})$, with chemical potential $\mu$, observed in a reference frame rotating with angular velocity $\Omega$, obeys the dimensionless GPE
\begin{equation}
\label{eq:GPEindiv}
(i-\gamma)\frac{d\psi}{dt} = -\nabla^2\psi -(\mu-V-|\psi|^2)\psi-\Omega \hat{L}_z\psi~,
\end{equation}
where the sound speed is $c_s = (n_0 g/m)^{1/2}$, the healing length (defined as the characteristic length-scale) is $\xi = \hbar/(m n_0 g)^{1/2}$, the characteristic time-scale is $\hbar/(n_0 g)$, the energy scale is $n_0 g$,  $g$ is the self-interaction strength, $m$ is the mass of each boson (twice the neutron mass), and $n_0$ is the particle density.  $\hat{L}_z=-i\partial/\partial\theta$ is the antisymmetric angular momentum operator.

The presence of a thermal cloud also gives rise to a mutual friction force, which is self-consistently modelled in hydrodynamic treatments using the HVBK equations \cite{Hills:1977}, or alternatively by solving Eq.~(1) simultaneously with a similar equation describing the quantum mechanics of the excited states \cite{Penckwitt:2002p1045}.  Practically, the consequence of neglecting mutual friction in our treatment is that we cannot expect to correctly capture behaviour stemming from the drag between the inviscid and viscous components.

We also note that Eq.~(1) applies to weakly interacting systems such as dilute atomic Bose gases, in which short-range, two-body interactions dominate \cite{Roberts:2001LNP}.  For a more strongly correlated superfluid, which experiences drag from the non-condensed portion, a nonlocal interaction potential is more appropriate, which leads to a phonon-roton spectrum \cite{Pomeau:1993PRL}.  We are therefore cautious in relating our results to superfluid flow in laboratory experiments and in neutron stars.

Equation~(2) demonstrates clearly that the system is governed by four variables:  $\gamma$, $\mu$, $V$ and $\Omega$.  Dissipation \citep{Choi:1998p9985,Tsubota:2002p11,Kasamatsu:2003p1051} is controlled by $\gamma$.  Experiments on damped oscillations in trapped atomic Bose-Einstein condensates performed by \cite{Mewes:1996p9989} and modelled by \cite{Choi:1998p9985} imply $\gamma \approx 0.03$.  In our simulations, we set $\gamma = 0.05$.  Its inclusion in Eq.~(\ref{eq:GPEindiv}) is equivalent to propagating the system in imaginary time, accelerating numerical convergence, as sound waves are quickly damped, and ensuring that $\psi$ can adjust in time to non-adiabatic changes in $V$ and $\Omega$.  Dissipation arises when atoms are exchanged between the thermal cloud and the condensate \citep{Jackson:2001p10058,Penckwitt:2002p1045,Jackson:2009p8818}, a process not modelled directly in this paper.  The dissipative term drives particle loss, as the normalisation of $\psi$ decays with time.  To correct for this artificial side effect, at each time step we advance $\mu$ according to a prescription used widely in the literature \citep{Kasamatsu:2003p1051}
\begin{equation}
\label{eq:deltamuindiv}
\mu(t+\Delta t) = \mu(t)+\frac{1}{\Delta t}\log\left[\frac{\int|\psi(t+\Delta t)|^2d^3x}{\int|\psi(t)|^2d^3x}\right]~.
\end{equation}  

Using a fourth-order Runge-Kutta algorithm in time, and a fourth-order finite difference scheme in space, we solve Eq.~(\ref{eq:GPEindiv}) in the co-rotating frame on a $100\times100$ square grid.  Unless stated otherwise, the time step is $\Delta t=0.005$ and the spatial grid resolution is $\Delta x=0.02$ in dimensionless units.

The trapping potential is circularly symmetric, with a radial profile defined by $V_{\rm{trap}}(r)=V_{\rm{max}}\left\lbrace 1+\tanh[2(r-R)]\right\rbrace$, with $V_{\rm{max}}=200$ for the experiments described in this paper.  A circularly symmetric potential alone does not nucleate vortices; a non-axisymmetric component (like the pinning grid) is needed to transfer angular momentum to the condensate.  Each site in the pinning grid, which is stationary in the corotating frame, takes the form
\begin{equation}
\label{eq:Vpinindiv}
V_{i,\rm{pin}} (\mathbf{r})= V_i\left[1+\tanh|\Delta(\mathbf{r}-\mathbf{R}_{i})|\right]~,
\end{equation}
where $\mathbf{R}_{i}$ is the position of the center of the spike, $V_0$ controls the pinning strength, and $\Delta = 4$ parametrises the width, with $V=V_{\rm{trap}}+V_{i,\rm{pin}}$.

Given the unitary evolution inherent in Eq.~(\ref{eq:GPEindiv}), it is not strictly correct to describe vortex unpinning as a binary on-off process. Semi-classical treatments \citep{Hakonen:1998p10787} describe the unpinning rate of a vortex, pinned by a potential of strength $V_i$ by the Arrhenius formula $\Gamma =\Gamma_0 e^{-V_{i,\rm{pin}}/(kT)}~\rm{s}^{-1}$, where $\Gamma_0$ is the zero-point attack rate against the pinning barrier, and $T$ is the temperature.  

In this paper we track the evolution of $|\psi|^2$, the expectation value of the condensate density.  This mean-field approach describes the average behaviour of a vortex, without quantum fluctuations.

\section{\label{sec:shear}Shear driven unpinning}

In this section we study how a vortex unpins when the surrounding condensate flows past it in bulk, in situations where the relative motion is forced by \textit{global, mean-field shear} (between the bulk condensate and the pinning array) rather than local shear induced by neighbouring vortices.  In analogy with laminar flow past a whirlpool, a vortex experiences a Magnus force when it moves relative to the ambient condensate \citep{Donnelly}.  The force is transverse to the direction of differential motion; for a pinned vortex line with velocity $\mathbf{v}$ moving through a condensate with velocity $\mathbf{v}_{\rm{S}}$, the Magnus force per unit length is
\begin{eqnarray}
 \mathbf{F}_{\rm{M}}&=&\rho\bm{\kappa}\times(\mathbf{v}-\mathbf{v}_{\rm{S}})~,
\label{eq:Magnusindiv}
\end{eqnarray}
where $\kappa$ is the circulation vector (out of the plane in two dimensions, magnitude $4\pi$ in dimensionless units).  For a vortex pinned at radius $b$ to a container rotating with angular velocity $\Omega$, in a bulk condensate approximating rigid-body rotation with angular velocity $\Omega_{S}$, we have $|\mathbf{F}_{\rm{M}}|\approx\rho\kappa b(\Omega-\Omega_{\rm{S}})$.  An important corollary of Eq.~(\ref{eq:Magnusindiv}) is that, in the absence of pinning, a vortex revolves around the centre of the trap with the condensate ($\mathbf{v}=\mathbf{v}_{\rm{S}}$).  We refer to $\mathbf{v}=\mathbf{v}_{\rm{S}}$ in Sec.~\ref{subsec:ballistic} as \emph{ballistic, unforced} motion.  In order to relate $F_{\rm{M}}$ to the pinning potential that appears in Eq.~(\ref{eq:GPEindiv}), we must specify a characteristic length-scale, $\xi$, over which the pinning potential acts, with $F_{\rm{M}} \leq V_i/\xi^2$ for the vortex to remain pinned.

The initial conditions of the numerical experiment are created by imposing a rectangular pinning array to nucleate and pin vortices.  The trapping potential (i.e. the container) is stationary  in the rotating frame and takes the form
\begin{eqnarray}
\label{eq:potential}
V_{\rm{trap}}(r) &=& V_{\rm{max}}\left[ 1+\tanh(2r)\right]~,
\end{eqnarray}
where $V_{\rm{max}}$ (= 200 in all simulations described in this paper) defines the maximum potential and $R$ is the cylindrical radius of the container\footnote{$V_{\rm{trap}}$ at the edge of the trap cannot be so steep that it is unresolved by the simulation grid; $V_{\rm{max}}$ must be large enough to stop particles leaking out of the trap.}.  An axisymmetric potential like Eq.~(\ref{eq:potential}) does not nucleate vortices; a non-axisymmetric component, for example a pinning grid, is essential to catalyse the formation of a vortex lattice.
Once a vortex array is nucleated, all but one of the pinning sites are removed instantaneously.  In response, all but one of the vortices move to new positions in a modified (to account for the still-pinned vortex) Abrikosov lattice; one vortex remains pinned, as shown in the left contour plot in the top row of Fig.~\ref{fig:diff_latt}.    Finally, the angular velocity of the container is decreased, causing the unpinned vortices to move radially outward and eventually annihilate at the container wall.  

In these experiments, we deliberately ignore the complications arising from multiple vortices embedded in a pinning grid.  In particular, in the absence of other vortices, the bulk condensate velocity exactly equals the self-induced velocity field, which does not contribute to the Magnus force.

The aim of this experiment is to track changes in the energy of the system when a vortex unpins, as a function of the strength of the potential that pins the vortex.  We are also interested in how and where the energy released during unpinning is transported.  To these ends, we want to start with a clean initial state with one off-axis vortex.  The total energy per unit vertical length of the condensate is calculated using
\begin{equation}
E = \int d^2 x\left( \psi^{\ast}\Omega \hat{L}_z\psi+|\nabla\psi|^2+V|\psi|^2+\frac{1}{2}|\psi|^4\right)~,
\end{equation}
Chiefly, we track changes in the lattice energy, $E_{\rm{latt}}=\int V|\psi|^2 d^2x$, and the kinetic energy, $E_{\rm{kin}}=\int|\nabla\psi|^2d^2x$, as a vortex unpins and moves.  

To unpin the sole pinned vortex in the initial state, we ramp the angular velocity of the container ($\dot{\Omega} = -10^{-3}$).  Negative values of $\Omega$ indicate rotation in the opposite sense to the flow induced by the vortex.  In Fig.~\ref{fig:diff_latt} and \ref{fig:diff_kin}, we present three cases:  $V_i=1.5V_0$, $2.0V_0$, and $2.5V_0$ (dotted, solid and dashed curves respectively; $V_0 = 60$).  The top panel of Fig.~\ref{fig:diff_latt} shows three snapshots of the condensate density $|\psi|^2$:  before the vortex unpins (left), as it moves towards the edge (centre), and as it annihilates against the wall (right).  The range of densities indicated by the colour scale corresponds to $\sim 20\%$ of the full range, to emphasise the low-amplitude sound waves emitted by the moving and annihilating vortex in the centre and right snapshots respectively.  The scale runs from low (dark) to high (light) density. The curves shown in the top panel of Fig.~\ref{fig:diff_kin} graph $E_{\rm{latt}}$ as a function of the angular velocity of the container.  For $V_i=2.0V_0$, at $\Omega\approx -0.661$ the vortex begins to unpin and $E_{\rm{latt}}$ simultaneously increases, as $|\psi|^2$ increases at the pinning site where previously there was a density minimum.  We observe oscillations in $E_{\rm{latt}}$ for $-0.661<\Omega <-0.670$ as the sound waves emitted by the moving vortex pass over the pinning site and/or collide with the wall (see ripples emanating from the left side of the right contour plot).  In all three cases, there is a net decrease in $E_{\rm{latt}}$ when the vortex unpins and annihilates, demonstrating that the pinned state is a local energy minimum (\emph{i.e.} metastable), instead of a true ground state.  The threshold shear for unpinning scales as $\Delta\Omega_{\rm{crit}}\approx -0.33V_i$.

In Fig.~\ref{fig:diff_kin} we plot the contribution to $E_{\rm{kin}}$ from a unit disk at the origin.  The $\sim 1\%$ variation in the pre-unpinning value of $E_{\rm{kin}}$ between the three different pinning scenarios arises because a strongly pinned vortex is dragged through the condensate faster than a weakly pinned vortex (\emph{i.e.} $\Delta\Omega\propto V_i$).  Compared to the change in $E_{\rm{kin}}$ when the vortex unpins ($\sim 40\%$), however, this difference is negligible.  Once the vortex unpins, it moves radially outward, away from the unit disk; the vortex-induced velocity at the origin is inversely proportional to the distance to the vortex.  The jump in $E_{\rm{kin}}$ ($\Delta E_{\rm{kin}}/E_{\rm{kin}}\approx 1.27$) when the vortex finally annihilates against the wall is followed by acoustic oscillations ($\Delta E_{\rm{kin}}/E_{\rm{kin}}\approx 0.5$), which are visible in the right contour plot in the top row of Fig.~\ref{fig:diff_latt}.

In the context of neutron star glitches, the results of many experiments like these confirm that a vortex always unpins at some $\Delta\Omega_{\rm{crit}}$ in a global shear and always changes $E_{\rm{latt}}$ by some \textit{fixed} amount.  By contrast, pulsar glitches have a power-law spread of event energies and trigger waiting times (see Sec.~\ref{sec:intro}).  In the following sections, we investigate how global shear unpinning can trigger subsequent multiple unpinnings through acoustic radiation and vortex proximity effects, producing overall stochasticity through cooperative processes.  The effects of acoustic and proximity knock on are studied below in the context of a single pair of vortices, where it is easier to track what is happening than in a vortex array.  An abridged discussion of the many-vortex behaviour is presented for completeness in the Appendix.

\section{\label{sec:acoustic}Acoustic knock on}

One possible trigger for an unpinning avalanche  is the emission of sound waves by a moving vortex.  In this section, we demonstrate that sound waves can indeed unpin vortices. In the first instance, we unpin a vortex using artificially generated sound waves.  We then demonstrate that sound waves generated by a spontaneous repinning event are sufficient to unpin a nearby vortex. 

\subsection{\label{subsec:art_sound}Artificially generated sound waves}

\begin{figure*}
\begin{center}
\hspace{1.7cm}\includegraphics[scale=0.65,angle=90]{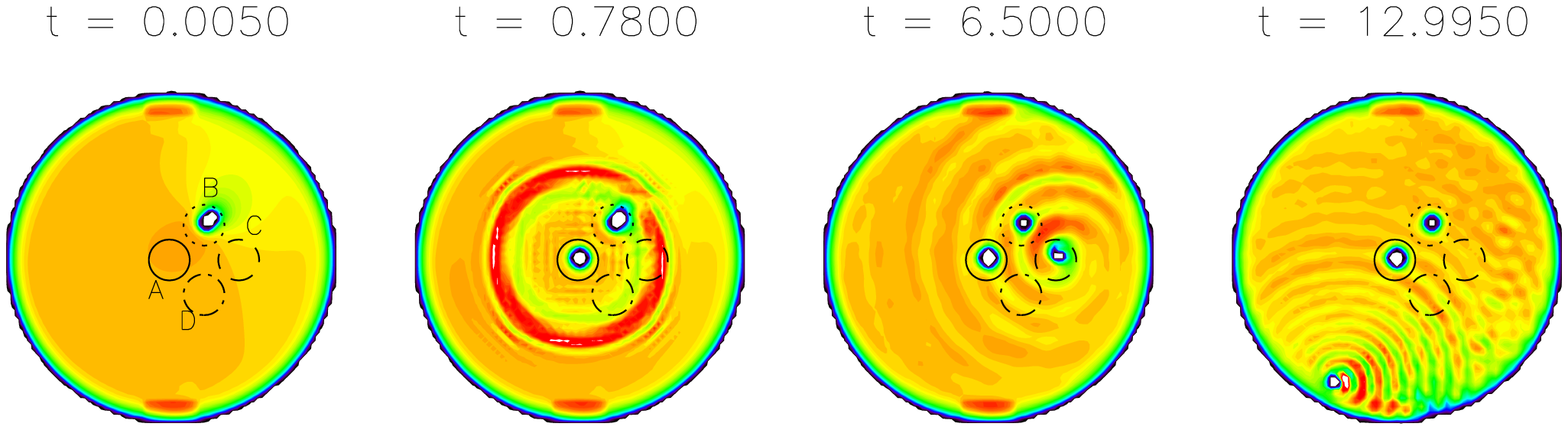}
\vspace{-11cm}\\
\includegraphics[scale=0.63,angle=90]{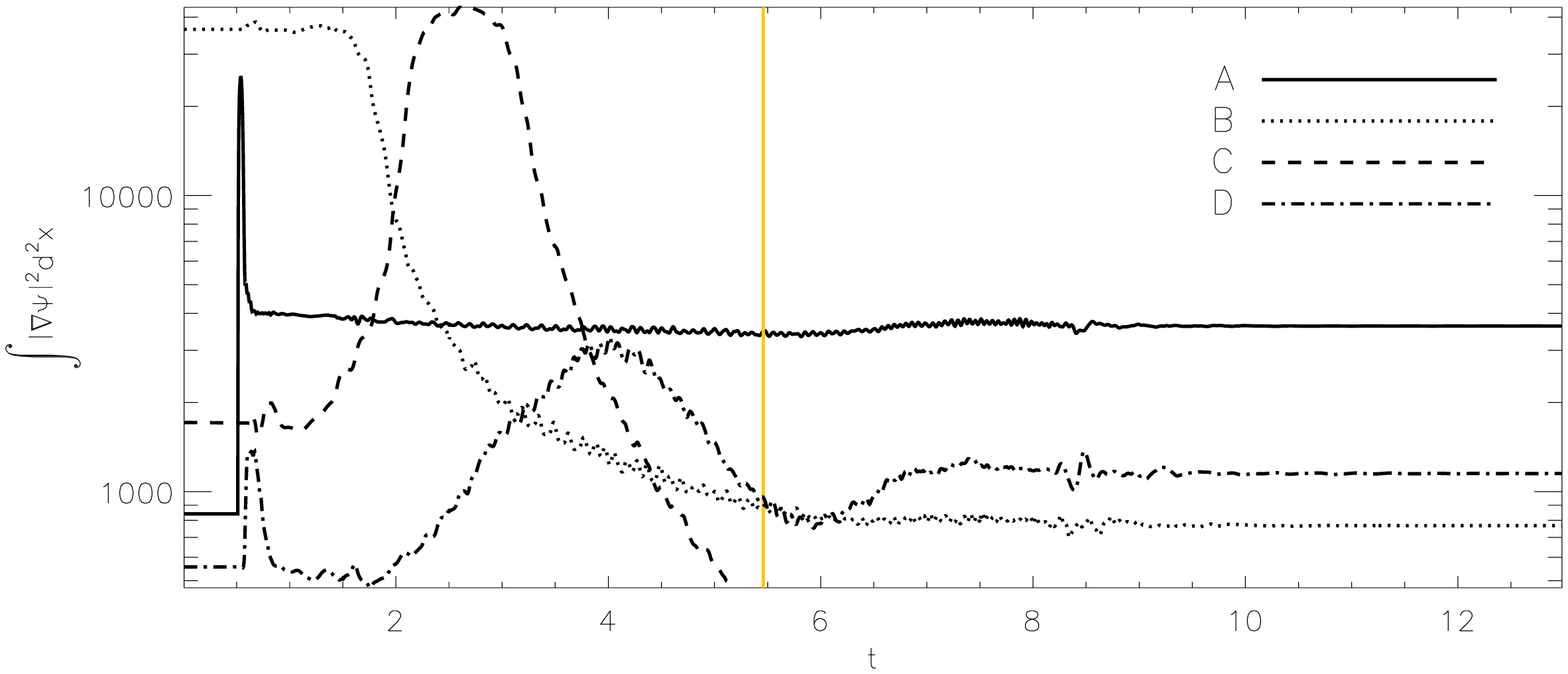}
\end{center}
\caption{\emph{Top}:  Snapshots of the condensate density $|\psi|^2$ (colour; dark/light represents low/high density) at times $t=0.005$, 0.780, 6.500 and 12.995 (left to right).  A vortex initially pinned (with strength $2V_0$) at (1.7,1.7) is unpinned by an acoustic pulse launched by an impulsive spike in the potential of strength $4V_0$ at (0.0,0.0) at $t=0.5$. \emph{Bottom}: Kinetic energy $E_{\rm{kin}}$ integrated within the four unit disks in the top panels, which we label to (clockwise from left) as regions A--D.  Note that the vertical axis in the bottom panel is logarithmic.  The vertical grey line at $t=5.49$ marks the time of the snapshot in Fig.~\ref{fig:sound_unpin_cross} below.  Simulation parameters:  $R=12.5$, $\Omega=-0.65$, $V_{\rm{trap}}=200$ and $\gamma =0.02$. }
\label{fig:sound_unpin_snap}
\end{figure*} 
\begin{figure*}
\begin{center}
\includegraphics[scale=0.625,angle=90]{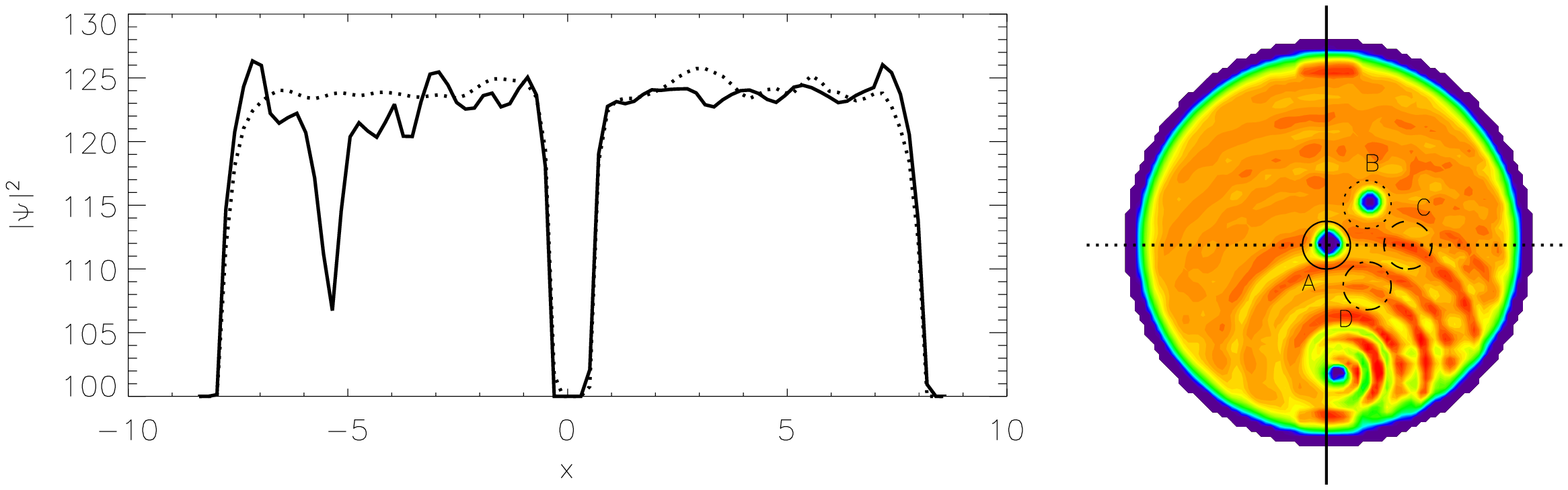}
\end{center}
\caption{\emph{Left}: Cross-sections of the condensate density $|\psi|^2$  taken vertically (solid curve) and horizontally (dotted curve) through a snapshot at $t=5.49$ from the simulation in Fig.~\ref{fig:sound_unpin_snap}. \emph{Right}:  Contours of $|\psi|^2$ (colour; dark/light represents low/high density).}
\label{fig:sound_unpin_cross}
\end{figure*} 
 
The initial setup of the numerical experiments described in this subsection is similar to the setup in Fig.~\ref{fig:diff_latt}.  A vortex is pinned at (1.7,1.7) (region B in the left colour plot in Fig.~\ref{fig:sound_unpin_snap}) by a pinning site with $V_i=2.0V_0$.  The container and pinning site rotate at $\Omega=-0.65$ (in the opposite sense to the flow induced by the vortex), so that the metastable pinned state is `stressed' as it is not the energy minimum.  At $t=0.5$, a pinning spike [centred at (0,0) in region A] with functional form given by Eq.~(\ref{eq:Vpinindiv}) is instantaneously raised to $V_i=4.0V_0$.  This launches a circular acoustic pulse with wavelength $\approx 5$ and phase speed $c_{\rm{s}}\approx 2$.  When the pulse passes over the pinned vortex, it successfully unpins it.  The vortex then moves outward along a spiral trajectory. 

In the bottom panel of Fig.~\ref{fig:sound_unpin_snap}, we plot the partial contributions to $E_{\rm{kin}}$ from four non-overlapping regions: unit disks centred at (0.0,0.0), (1.7,1.7), (1.7,$-1.7$) and (0.0,3.4) (hereafter regions A--D respectively, graphed as solid, dotted, dashed and dot-dashed curves respectively).  Regions A and B sit over the origin of the acoustic pulse and the site of the pinned vortex respectively.    When the pulse is launched at $t=0.5$, $E_{\rm{kin}}$ jumps in region A (solid curve).  As the acoustic front moves radially outwards, it passes through regions B and C approximately 0.125 time units later, accompanied by jumps in $E_{\rm{kin}}$.  The progress of the unpinned vortex towards the wall of the container can also be tracked by monitoring $E_{\rm{kin}}$, which decreases in region B as the vortex moves away.  The vortex inhabits region C in the interval $2.5\lesssim t\lesssim 3.0$, accompanied by a broad peak in the dashed curve.  The annihilation of the vortex against the wall of the container registers in all four regions as a $\sim 10\%$ oscillation in $E_{\rm{kin}}$.

The colour plot of condensate density in the right panel of Fig.~\ref{fig:sound_unpin_cross} is a snapshot of $|\psi|^2$ taken at $t=5.475$, when the vortex is travelling towards the wall. The sound waves are accentuated by restricting the plotted contours to $100\leq |\psi|^2\leq 150$.  The curves in the left panel of Fig.~\ref{fig:sound_unpin_cross} are cross-sections along vertical (solid curve) and horizontal (dotted curve) diameters at $t=5.475$.  The moving vortex generates density fluctuations with peak-to-peak amplitude $\lesssim 2\%$ of the mean density.

\subsection{\label{subsec:vortex_sound}Sound waves from spontaneous vortex repinning}
\begin{figure*}
\begin{center}
\hspace{1.7cm}\includegraphics[scale=0.65,angle=90]{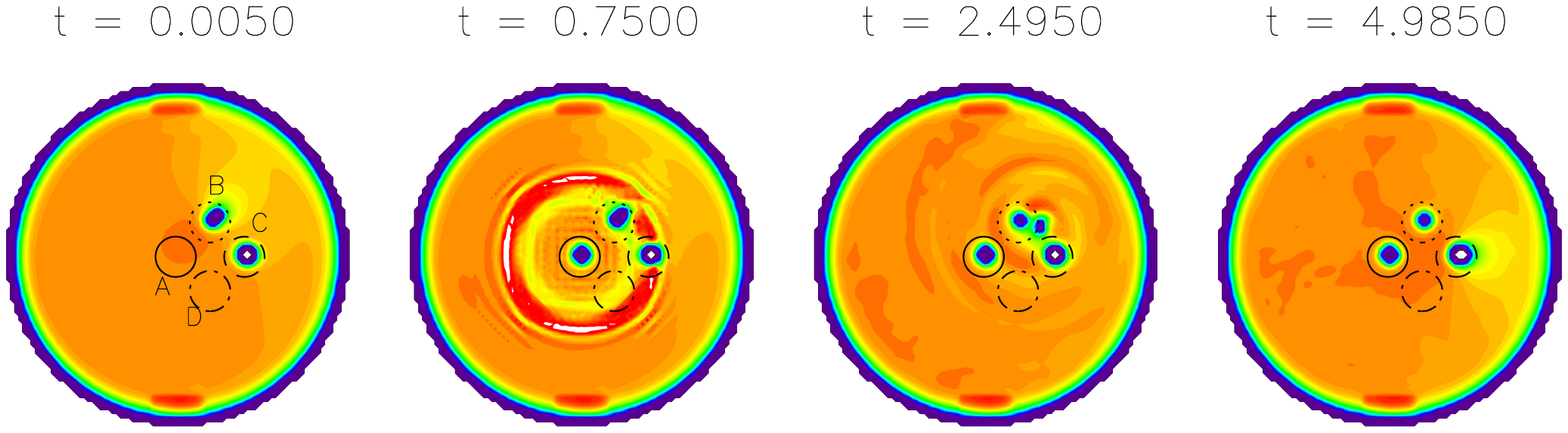}
\vspace{-11cm}\\
\includegraphics[scale=0.63,angle=90]{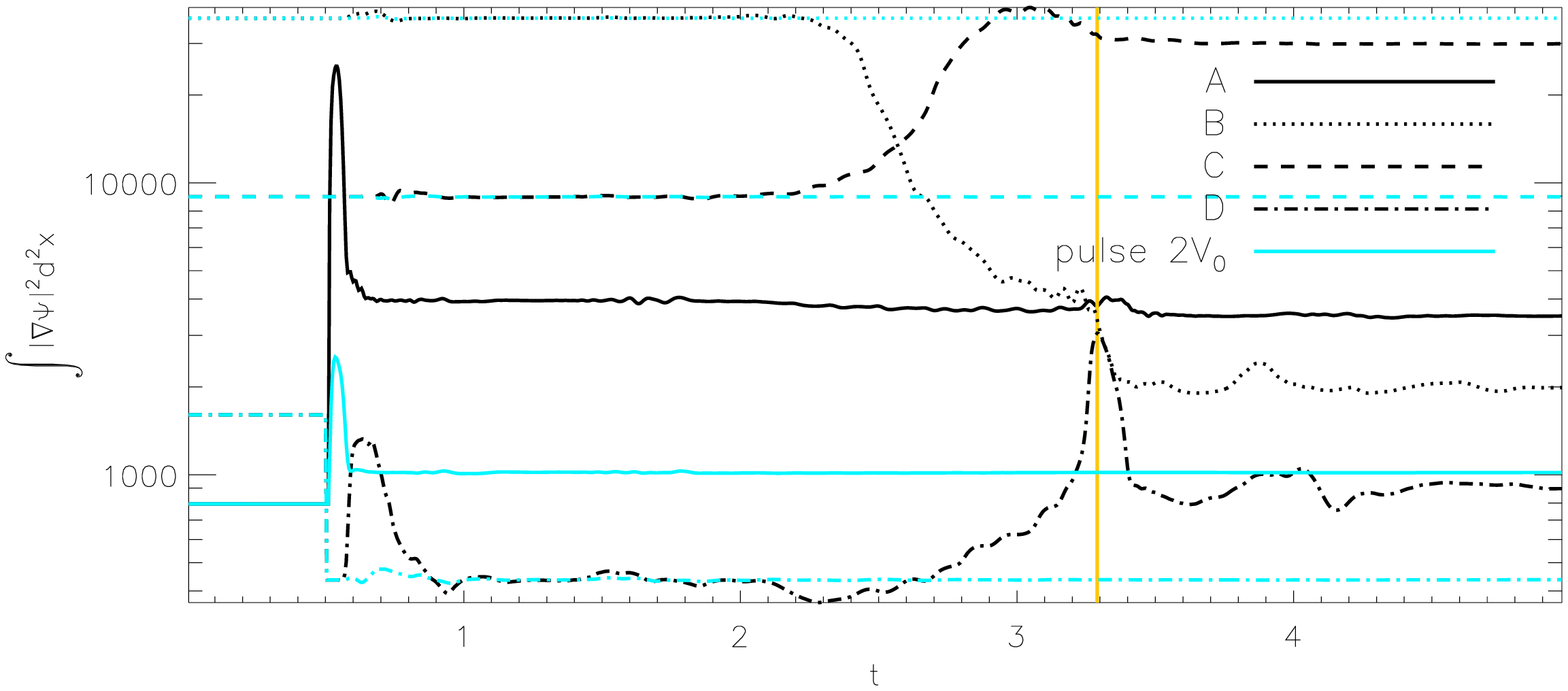}
\end{center}
\caption{\emph{Top}: Snapshots of the condensate density $|\psi|^2$ (colour; dark/light represents low/high density) at times $t=0.005$, 0.750, 2.495 and 4.985 (left to right respectively).  A vortex is that is initially pinned (with strength $1.97V_0$) in region B is unpinned by an impulsive acoustic pulse of strength $4V_0$, launched in region A, and then repins at a pinning site in region C ($V_i=6V_0$).  The snapshots are taken before the initial sound pulse (left), as the wavefront from the sound pulse impacts on the pinned vortex (second from left), as the vortex travels between the two pinning sites (third from left), and once the vortex has repinned at C (right).  \emph{Bottom}: Kinetic energy $E_{\rm{kin}}$ integrated within the four unit disks in the top panels, labelled regions A--D.  Note that the vertical axis in the bottom panel is logarithmic.  The vertical grey line at $t=3.29$ marks the time of the snapshot in Fig.~\ref{fig:sound_repin_cross} below.  Simulation parameters:  $R=12.5$, $\Omega=-0.65$, $V_{\rm{trap}}=200$ and $\gamma =0.02$. The grey curves correspond to a control experiment, in which the initial pulse (amplitude $2V_0$) is too weak to unpin the vortex.}
\label{fig:sound_repin_snap}
\end{figure*} 
\begin{figure*}
\begin{center}
\includegraphics[scale=0.625,angle=90]{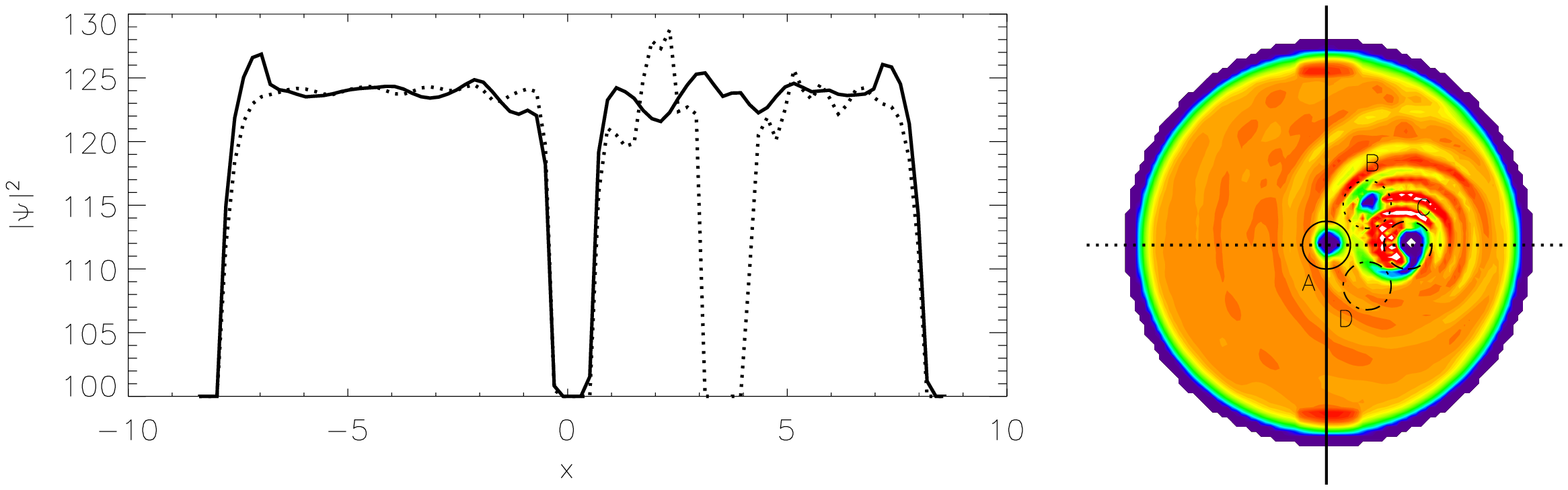}
\end{center}
\caption{\emph{Left}: Cross-sections of the condensate density $|\psi|^2$ taken vertically (solid curve) and horizontally (dotted curve) through a snapshot at $t=3.29$ from the simulation in Fig.~\ref{fig:sound_repin_snap}. \emph{Right}:  Contours of $|\psi|^2$ (colour; dark/light represents low/high density).}
\label{fig:sound_repin_cross}
\end{figure*} 

Let us now repeat the experiment described in Sec.~\ref{subsec:art_sound} by adding a pinning site in the prospective path of the unpinned vortex, to give it a place to repin.  The basic setup is discernible in the first of four contour plots at the top of Fig.~\ref{fig:sound_repin_snap}.  Once again we start with a pinned vortex ($V_i=1.97V_0$) in region B and a sound pulse launched from region A.  The additional pinning site sits inside region C ($V_i=6V_0$).  We are obliged to choose $V_{i,C}>V_{i,B}$ for two reasons:  (i) the Magnus force is linearly proportional to $r$; and (ii) if a vortex represents a local (not global) minimum in $E$, then once the vortex unpins it exits the condensate by annihilating against the wall.  In other words, it is more difficult to repin a moving vortex than it is to keep a stationary vortex pinned \citep{Link:2009p9063}.

The four contour plots in Fig.~\ref{fig:sound_repin_snap} are snapshots of the condensate density at times chosen to illustrate the process of vortex unpinning and repinning.  The second snapshot is taken 0.25 time units after the sound pulse is launched, when the wave front impacts on the pinned vortex in region B.  The vortex begins to unpin at $t=2.3$.  The delay ($\Delta t\approx 1.5$) between the sound pulse arriving and the vortex leaving the pinning site is routinely observed in these simulations.  In the third snapshot, the vortex is halfway between the two pinning sites.  In the final snapshot, the vortex pins to the pinning site in region C.  Repinning is accompanied by the generation of sound waves, which modulate $|\psi|^2$ by up to $15\%$ (cf. 2\% in Fig.~\ref{fig:sound_unpin_cross}).  This occurs as the vortex spirals into the pinning centre \cite{Sedrakian:1995MNRAS} in a manner akin to a basketball circling the rim of a hoop.  Repinning occurs $\sim 7.5$ times faster than unpinning, hence the stronger burst of radiation.

Having already observed the unpinning event in the previous experiment (Sec.~\ref{subsec:art_sound}), this study is primarily concerned with the emission resulting from repinning.  Once again we track changes in $E_{\rm{kin}}$ in four non-overlapping regions A--D as functions of time, as shown in the bottom panel of Fig.~\ref{fig:sound_repin_snap}.  The grey curves correspond to a control experiment, in which the initial pulse (amplitude $2V_0$) is too weak to unpin the vortex.  The vertical line indicates the final stage of repinning, when the snapshot in Fig.~\ref{fig:sound_repin_cross} is taken. In this snapshot, a \emph{spiral tail} drags behind the newly-pinned vortex; acoustic radiation from the repinning process is not axisymmetric.  The tail is recorded in $|\psi|^2$ in the left panel of Fig.~\ref{fig:sound_repin_cross} as a large fluctuation in the horizontal (dotted curve) but not the vertical (solid curve) cross-section.  Evidence for the tail is also found in the contributions to $E_{\rm{kin}}$ from regions A--D, e.g. a tripling in region D at $t=3.3$.

The kinetic energy tracks the condensate density flux through each unit disk, which in turn tracks distance to the vortex. For example, as the vortex moves closer to region D, $E_{\rm{kin}}$ in region D increases.  Conversely, there is a net decrease in region B, from where the vortex unpins and moves away.

\subsection{\label{subsec:sound_knock}Knock on}

\begin{figure*}
\begin{center}
\hspace{1.8cm}\includegraphics[scale=0.625,angle=90]{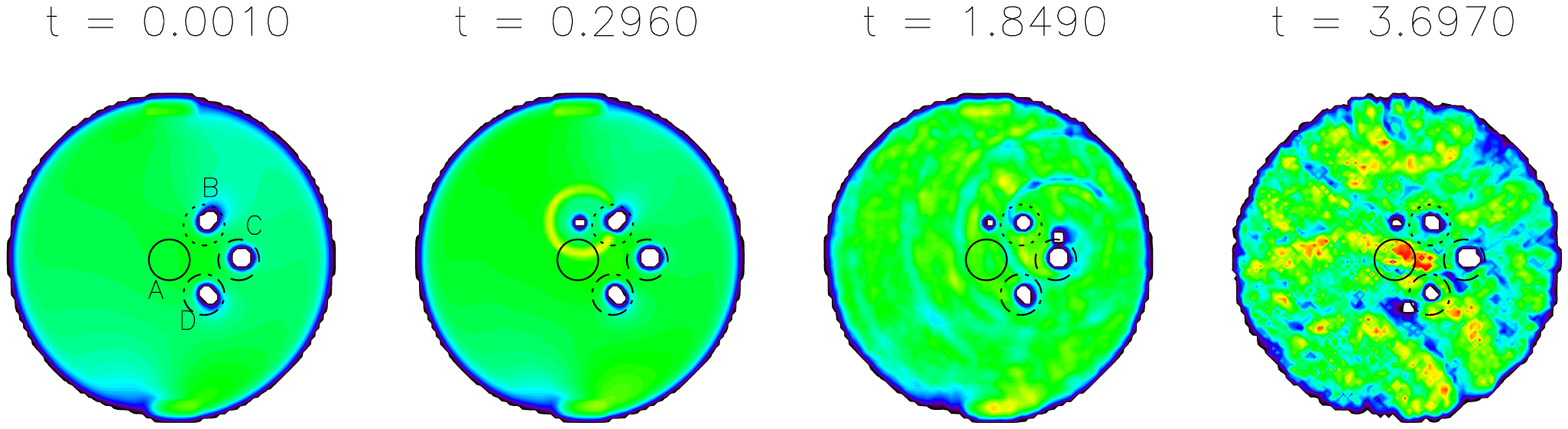}
\vspace{-11cm}\\
\includegraphics[scale=0.63,angle=90]{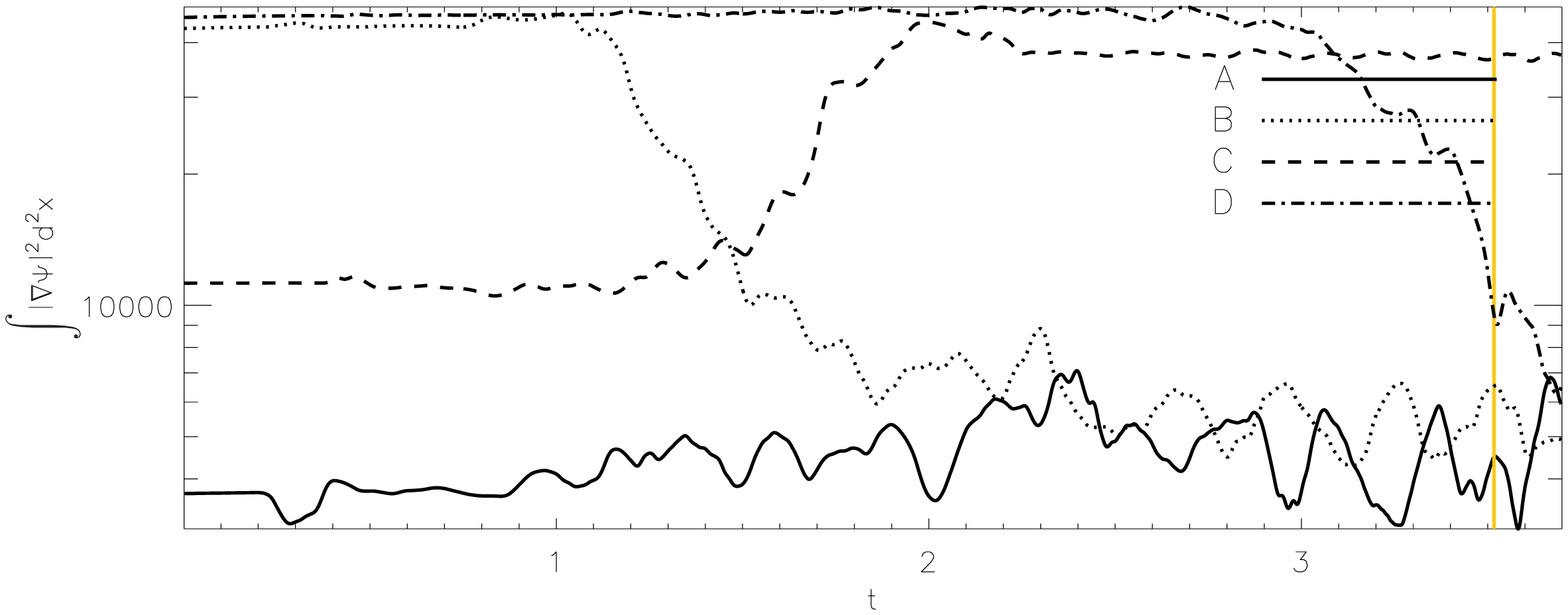}
\end{center}
\caption{\emph{Top}: Snapshots of the condensate density $|\psi|^2$ (colour; dark/light represents low/high density) at times $t=0.001$, 0.296, 1.849, 3.697.  A vortex that is initially pinned (with strength $2.9V_0$) in region B is unpinned by an impulsive acoustic pulse of strength $4V_0$, launched from (0.0,1.7), and then repins at a pinning site in region C ($V_i=6.75V_0$).  The sound waves emitted by the repinning vortex then strike and unpin a vortex pinned at D ($V_i=2.205V_0$).  The snapshots are taken before the initial sound pulse (left), as the wavefront from the sound pulse strikes the pinned vortex (second from left), as the vortex travels between the two pinning sites (third from left), and once the vortex in region D has unpinned and is travelling to the left of the container (right). \emph{Bottom}: Kinetic energy $E_{\rm{kin}}$ integrated within the four unit disks in the top panels, labelled regions A--D.  Note that the vertical axis in the bottom panel is logarithmic.  The vertical grey line at $t=3.29$ marks the time of the snapshot in Fig.~\ref{fig:sound_knock_cross} below.  Simulation parameters:  $R=12.5$, $\Omega=-0.65$, $V_{\rm{trap}}=200$ and $\gamma =0.0025$. }
\label{fig:sound_knock_snap}
\end{figure*} 

\begin{figure*}
\begin{center}
\includegraphics[scale=0.625,angle=90]{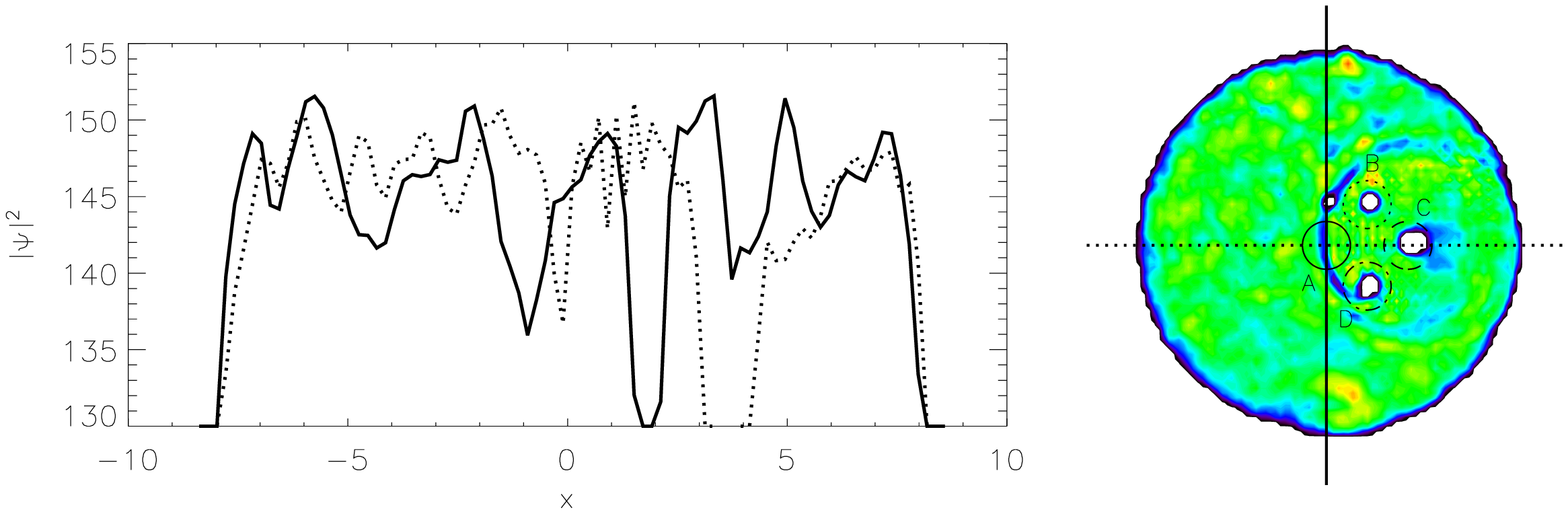}
\end{center}
\caption{\emph{Left}: Cross-section of the condensate density (shown in the contour plot at right) taken vertically (solid) and horizontally (dotted) through a snapshot at $t=3.51$ for the experiment shown in Fig.~\ref{fig:sound_knock_snap}, when the bottom vortex has unpinned and is travelling away from region D.  \emph{Right}: Contours of $|\psi|^2$ (colour; dark/light represents low/high density).}
\label{fig:sound_knock_cross}
\end{figure*} 

Finally, we construct an experiment that aims to demonstrate that the sound waves emitted by a moving vortex can unpin a nearby vortex.  In this example the initial pinning site has strength $V_i=2.9V_0$, the repinning site has strength $V_i=6.75V_0$, and there is a second vortex pinned in region D ($V_i=2.205V_0$).  The initial, artificially generated sound pulse is launched from (0.0,1.7) with amplitude $V_0$.  It is necessary to move the origin of the pulse away from region A so as not to simultaneously unpin the second vortex with the initial disturbance.  As in Sec.~\ref{subsec:vortex_sound}, the sound pulse unpins the vortex from region B, which then emits sound waves as it spirals in towards the pinning site in region C.  

When performed with the same level of dissipation as in the previous experiments (\emph{i.e.} $\gamma=0.02$), the radiation emitted during the basketball-in-a-hoop repinning process in region C is insufficient to unpin the vortex from region D.  When acoustic damping is reduced eight-fold ($\gamma=0.0025)$, however, the vortex unpins from region D, triggered by repinning in region C.  This is an important result for the understanding of the vortex dynamics leading to neutron star glitches, as it demonstrates the viability of unpinning avalanches arising from local acoustic triggers as well as global shear.

The four contour plots at the top of Fig.~\ref{fig:sound_knock_snap} are snapshots of the condensate density at representative times during the experiment in this section. The noisy ripples in Fig.~\ref{fig:sound_knock_snap} are not present in Fig.~\ref{fig:sound_repin_snap}, because acoustic disturbances are damped out by dissipation.  The contour plot in the right panel of Fig.~\ref{fig:sound_knock_cross} is a snapshot taken at $t=3.51$, when the second vortex is moving towards the left.  Cross-sections along vertical and horizontal cuts are represented in the left panel, from which it can be seen that $|\psi|^2$ fluctuations pervade the condensate and are consistently above $10\%$ of the mean (cf. $\sim 2\%$ from a moving vortex in Fig.~\ref{fig:sound_unpin_cross}).  Extensive experimentation, not reported here for brevity, confirms that unpinning can be caused by individual pulses and/or persistent \emph{acoustic noise}.  Although it is difficult to differentiate between the pulse emanating from the repinning site and the turbulent sea of sound waves fed by the original pulse and the subsequent vortex motion from B to C, the knock-on event at D likely results from a combination of both.

\subsection{\label{subsec:damping}Dissipation}
\begin{figure}
\includegraphics[scale=0.425]{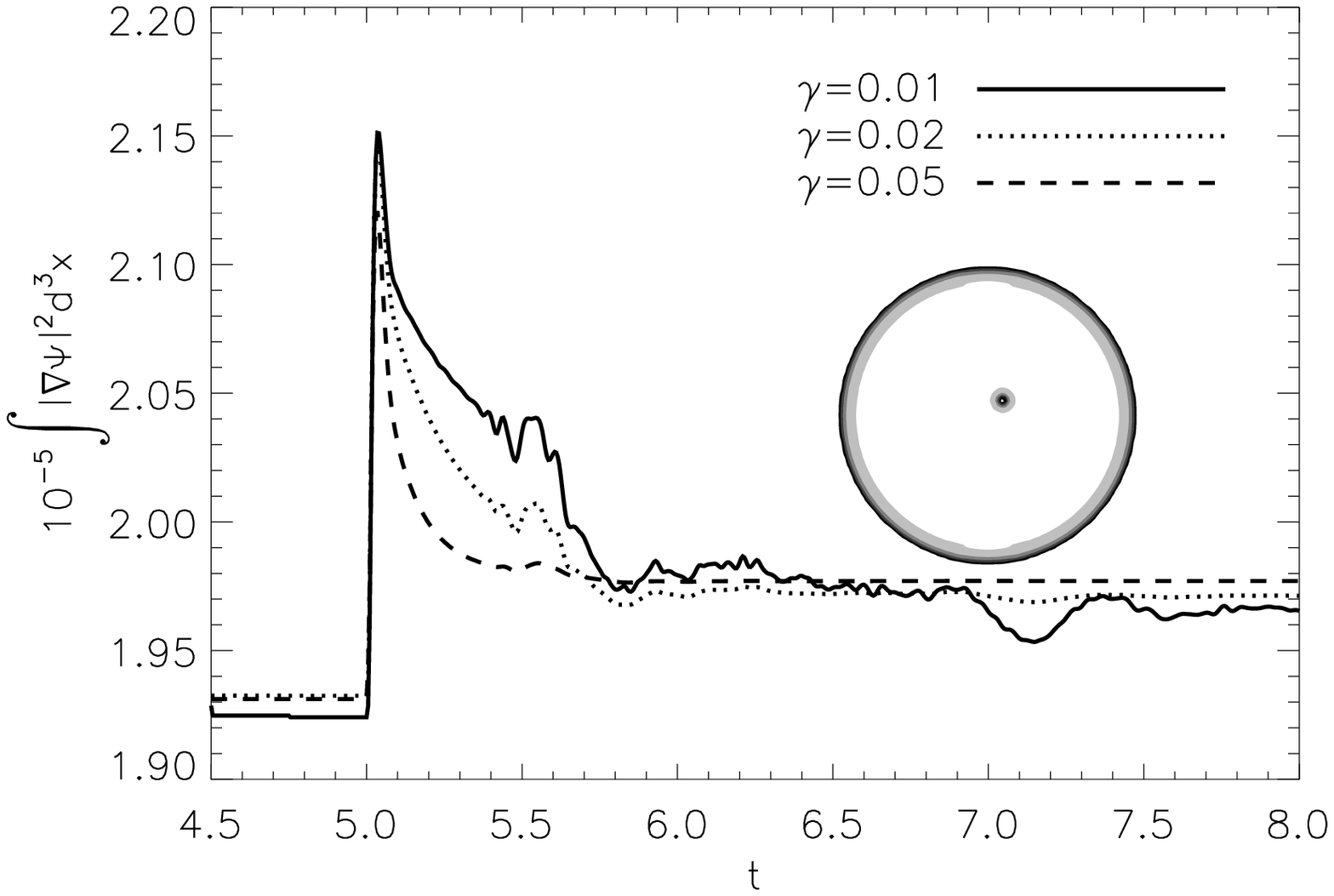}
\includegraphics[scale=0.425]{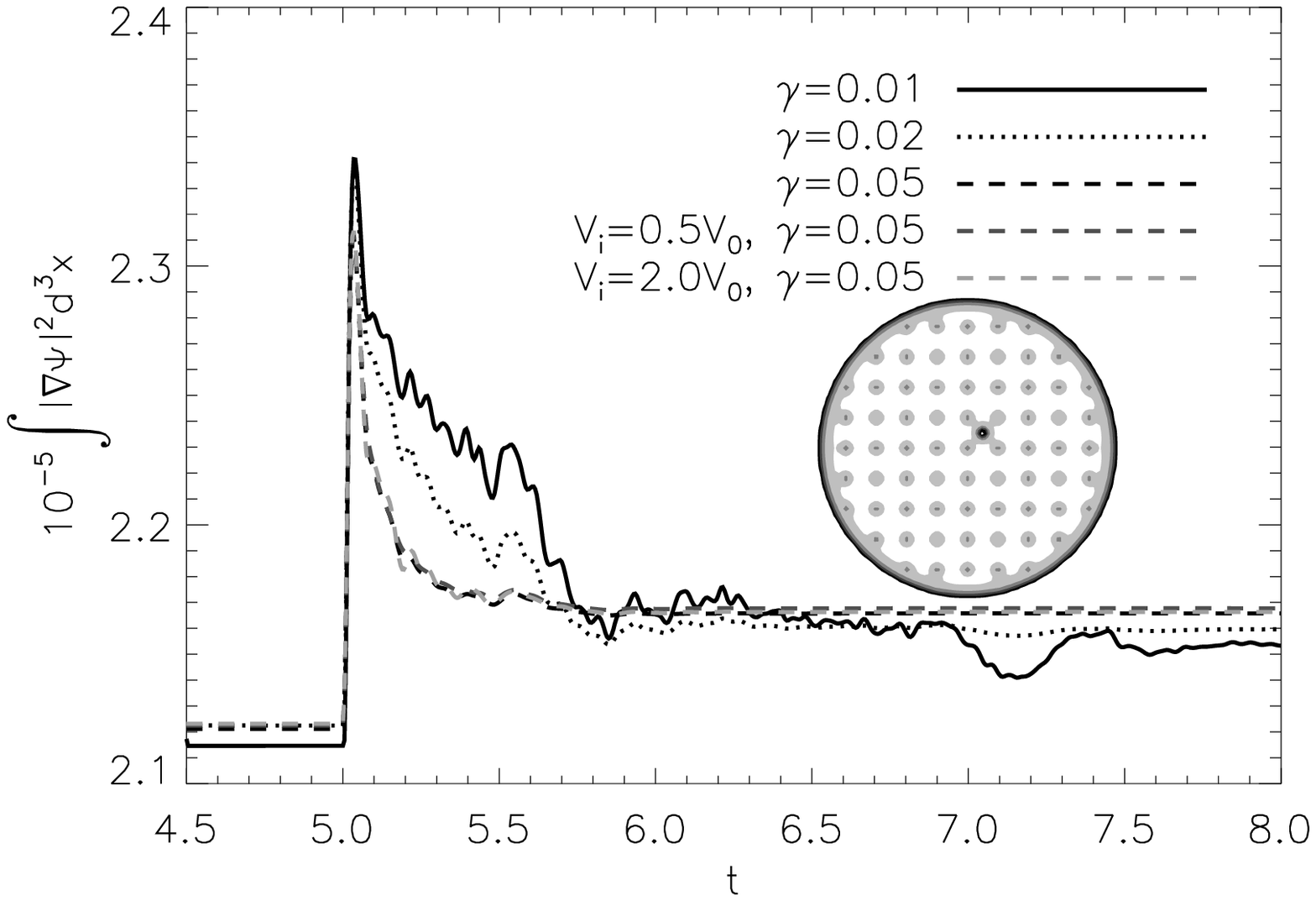}
\caption{\emph{Top}: Total kinetic energy $E_{\rm{kin}}$ as a function of time $t$ for a condensate in a stationary container of radius 8.5.  At $t=5.0$ a pulse of height $4V_0$ at (0.85,0.85) is generated.  The curves show three different levels of dissipation [$\gamma=0.01$, 0.02 and 0.05 (solid, dotted and dashed curves respectively)].  The inset shows a contour plot of the condensate density at $t=10$. \emph{Bottom}: As for top but with an $11\times11$ pinning grid [$V_i=0.5V_0$, $1.0V_0$ and $2.0V_0$ (grey, black and light grey curves respectively)].  The $0.5V_0$ and $2.0V_0$ curves have been shifted up and down respectively so that the initial value of $E_{\rm{kin}}$ agrees with that of the $1.0V_0$ curve.}
\label{fig:decay}
\end{figure} 

The numerical experiments described in Sec.~\ref{subsec:art_sound}--Sec.~\ref{subsec:sound_knock} are not conducted in the presence of a grid of pinning sites, cf. laboratory and astrophysical systems.  In order to evaluate the impact of introducing a pinning grid, in Fig.~\ref{fig:decay} we plot $E_{\rm{latt}}$ as a function of time for a pulse of strength $4V_0$ generated at (0.85,0.85) in a vortex-free condensate, at $t=5.0$ without (top panel) and with (bottom panel) a pinning grid.  We find that the time-scale on which the sound pulse decays is not affected by the pinning grid.  Conversely, when we repeat the experiment for three different levels of dissipation [$\gamma=0.01$, 0.02 and 0.05 (increasing dissipation)], we find that dissipation shortens the sound pulse decay time-scale irrespective of pinning.  We emphasise that unlike the damping of sound waves, vortex motion responds dramatically to a pinning grid; pinning sites either pin the vortex or re-route it around the site.    

The presence of other vortices is also pivotal in determining if a vortex unpins, and its ballistic trajectory.  Vortices in a uniform lattice are harder to unpin than isolated vortices for two reasons:  a vortex lattice lessens the differential rotation at each vortex; and, if another vortex is pinned radially beyond the vortex in question, vortex-vortex repulsion pushes a vortex back onto its pinning site.  These effects are explored further indirectly in the Appendix. 

\section{\label{sec:prox}Proximity knock on}
\begin{figure*}
\begin{center}
\includegraphics[scale=0.625,angle=90]{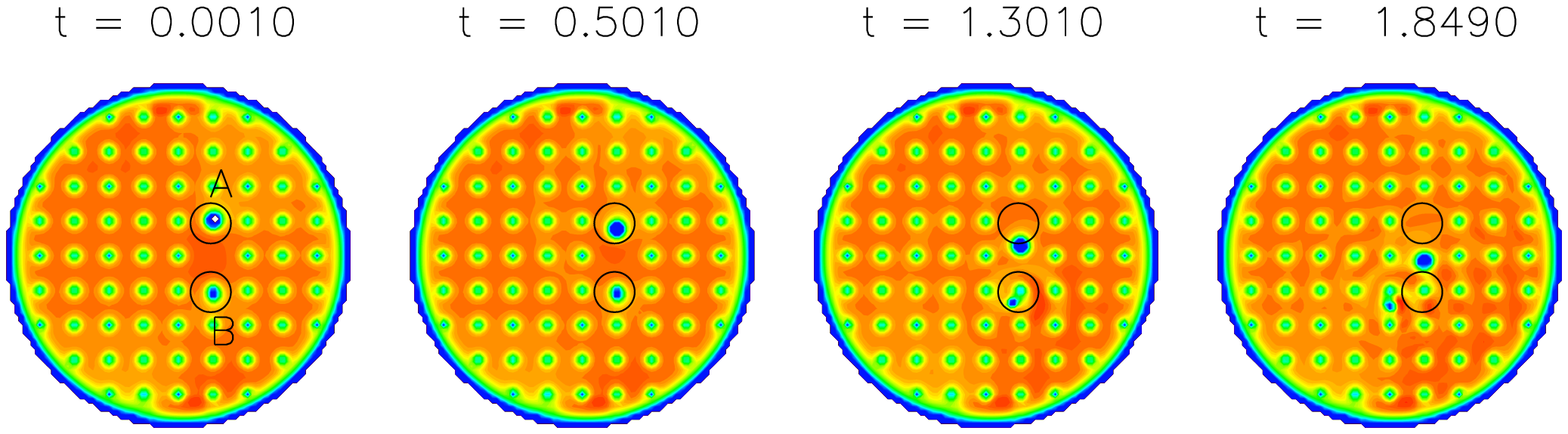}
\end{center}
\caption{Snapshots of the condensate density in a container of radius $R=8.5$ containing an $11\times11$ pinning grid, as a pinned vortex ($V_i=4.0V_0$), initially in region A, is dragged towards another, more weakly pinned vortex ($V_i=1.5V_0$) in region B.  The initial positions of the two vortices are indicated by black circles.  The container, including the pinning grid rotates with angular velocity $\Omega = -0.5$, in the opposite sense to the velocity field generated by the vortex.  The top vortex is dragged with unit dimensionless speed directly down towards the bottom vortex, unpinning it when the separation is 2.0.}
\label{fig:push}
\end{figure*} 
\begin{figure}
\begin{center}
\includegraphics[scale=0.375,angle=90]{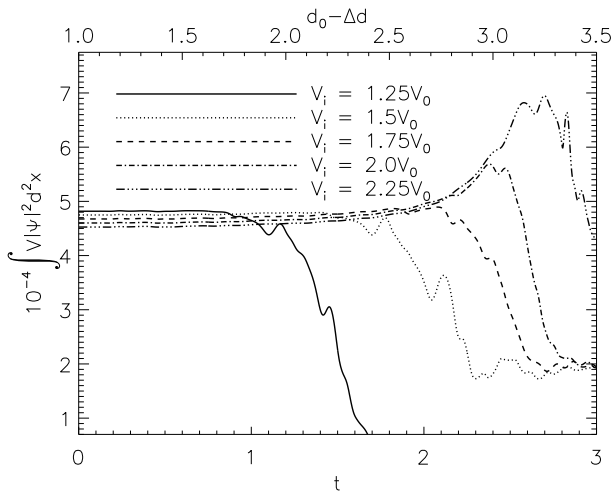}
\end{center}
\caption{Lattice energy, $E_{\rm{latt}}$ within a unit disk centred on the lower pinned vortex in Fig.~\ref{fig:push}, as a function of time (bottom axis) and vortex separation (top axis, $d_0$ is the initial vortex separation and $\Delta d$ is the distance traversed by the moving vortex).  The five curves correspond to five different pinning strengths for the lower vortex (see legend for $V_i$ values).  When the lower vortex unpins, $E_{\rm{latt}}$ drops sharply. }
\label{fig:Vpush}
\end{figure} 
\begin{figure}
\begin{center}
\includegraphics[scale=0.375,angle=90]{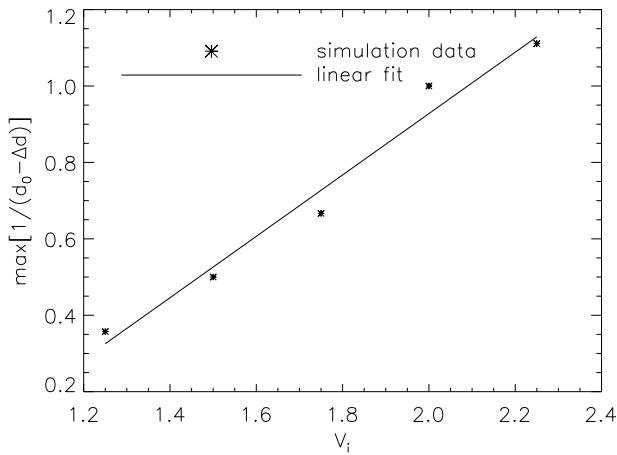}
\end{center}
\caption{Reciprocal of vortex separation $1/d_{\rm{unpin}}$ when the lower vortex in Figs~\ref{fig:push} and \ref{fig:Vpush} unpins, as a function of the potential it unpins from $V_i$.  The data points are overlaid by a linear regression, as predicted from Eq.~(\ref{eq:Magnusindiv}), with slope 0.825. }
\label{fig:dpush}
\end{figure} 

\begin{figure*}
\includegraphics[scale=0.6,angle=90]{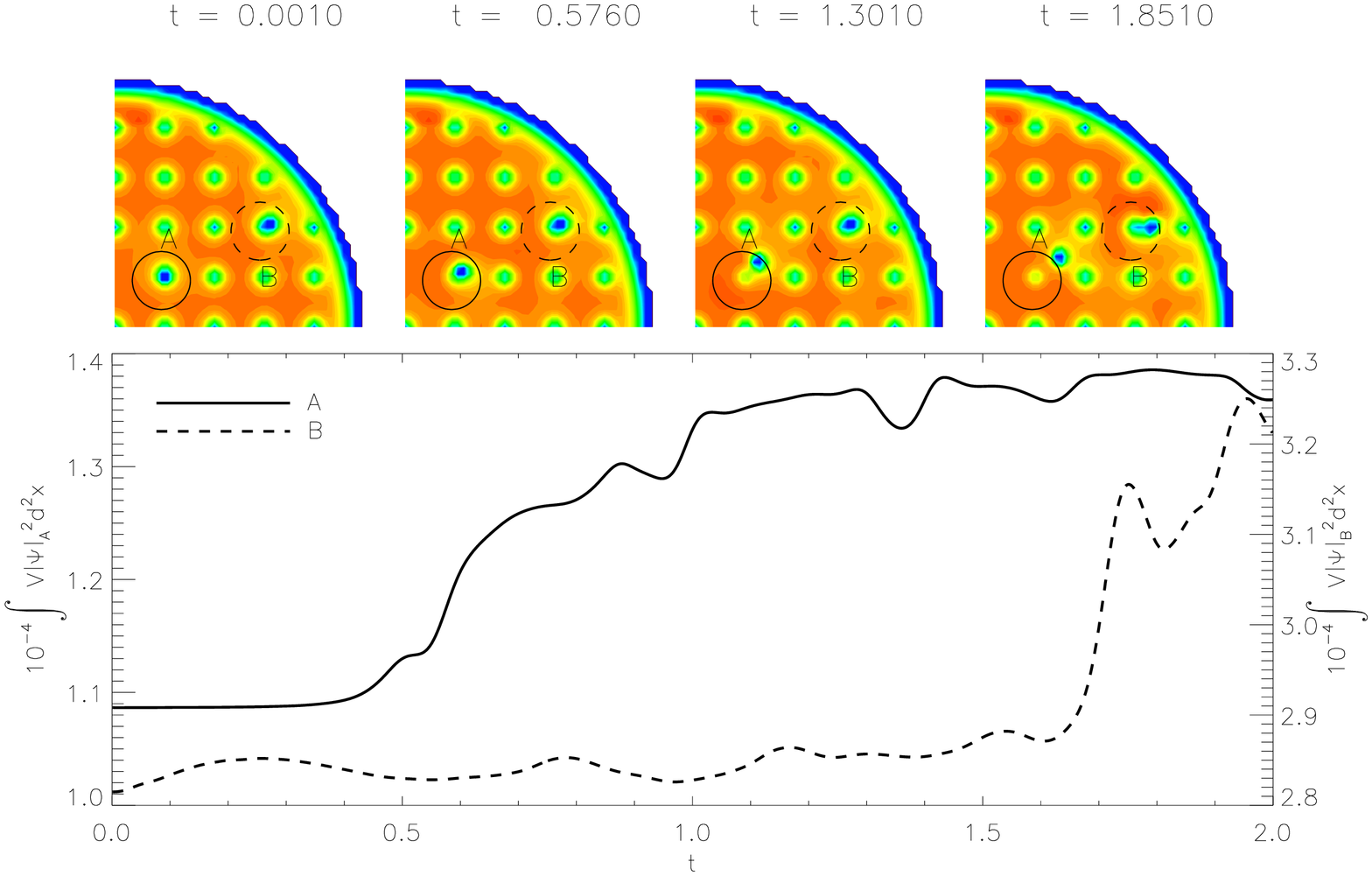}
\caption{\emph{Top}: A series of snapshots of the top right quadrant of the condensate density in a container of radius $R=8.5$ with a lattice of $11\times 11$ pinning sites of strength $V_0$.  Two pinned vortices are visible, as indicated by the solid (region A) and dashed (region B) circles (pinning strength $0.4V_0$ and $V_0$ respectively).  Four additional vortices are pinned in the other three quadrants. At $t=0$ the angular velocity of the container is reduced from $\Omega = 0.075$ to $\Omega = 0.05$. The vortex in region A unpins at $t\approx 0.4$ and travels towards region B.  When the intervortex separation is approximately 2.5 units, the vortex in region B unpins.  \emph{Bottom}: Contributions to $E_{\rm{latt}}$ (in dimensionless units) from region A (solid, plotted against the left-hand vertical axis) and region B (dashed, plotted against the right-hand vertical axis).  The sharp increase in each curve (at $t=0$ and $t=1.65$ respectively) marks the unpinning event.}
\label{fig:proxknock}
\end{figure*} 

In this section, we explore the possibility of knock on when an unpinned vortex approaches close to another while moving through a pinning grid before repinning.   When the distance between two vortices decreases from the Feynman spacing $d_0$ to $d_0-\Delta d$, the Magnus force per unit length $F_{\rm{M}}$ increases from $F_{\rm{M}_0}$ by 
\begin{equation}
\label{eq:magindiv}
\Delta F_{\rm{M}}=\frac{\rho\kappa^2}{2\pi}\frac{\Delta d}{(d_0-\Delta d)d_0}=\frac{F_{{\rm{M}}0}\Delta d}{d_0-\Delta d}~.
\end{equation}
We refer to the increase in Magnus force as the vortex proximity effect.  The vortex proximity effect unpins a vortex pinned by a potential $V_i$ if $F_{\rm{M}}+\Delta F_{\rm{M}}\geq \xi^{-2}V_i$, where $\xi$ is the condensate coherence length.  This allows vortices to remain marginally pinned while they are approximately in a Feynman lattice ($F_{\rm{M}}< \xi^{-2}V_i$), but they readily unpin in avalanches when a neighbour unpins and moves a little bit ($F_{\rm{M}}+\Delta F_{\rm{M}}> \xi^{-2}V_i$, with $\Delta F_{\rm{M}}\lesssim F_{\rm{M}0}$).  The coherence length and Coulomb lattice constant are comparable in a neutron star, creating an environment in which pinning by monovacancies and
intrinsic pinning in a polycrystalline structure are favourable \citep{Jones:1997p26}.   

The numerical experiments described below demonstrate the vortex proximity effect in two ways:  (i) by dragging a vortex towards a pinned vortex, and (ii) by allowing an unpinned vortex to move ballistically past other vortices as it travels freely towards the wall of the container.

\subsection{\label{subsec:proxdrag}Forced vortex motion}

Initial conditions are created by rotating a grid of pinning sites on which vortices nucleate and pin (see Sec.~\ref{sec:shear}).  All but two of the sites are then removed and the container is brought to a halt adiabatically, making the unpinned vortices annihilate at the wall.  All but one of the pinning sites are then reinstated, as depicted in the left contour plot in Fig.~\ref{fig:push}.  The missing site is deliberately situated between the two pinned vortices to let them be dragged towards each other.  The initial position of each vortex is indicated by a black circle; the upper (region A) and lower (region B) sites have $V_i=4V_0$ and $V_i=1.5V_0$ respectively.

We drag the upper vortex directly downwards by slowly moving its pinning site downwards with speed 1 (in dimensionless units, compared to a time step $\Delta t =0.001$).  The total distance traversed by the dragged vortex is $1.275$, which is $\sim 1.5d_{\rm{pin}}$ ($d_{\rm{pin}}$ is the spacing between pinning sites). We find that the lower vortex unpins (at a distance $d_{\rm{unpin}}$ from its nearest neighbour) when the approaching vortex gets within one pinning grid spacing $d_{\rm{unpin}}\approx d_{\rm{pin}}$.

We repeat this experiment for five values of $V_i$ in region B ranging from $1.25V_0$ to $2.25V_0$. Fig.~\ref{fig:Vpush} plots $E_{\rm{latt}}$ as a function of time for each experiment.  As $V_i$ increases, the unpinning event, indicated by the sharp drop in $E_{\rm{latt}}$, is delayed.  We also use movies of the $|\psi|^2$ to measure $d_{\rm{unpin}}$.  The top horizontal axis in Fig.~\ref{fig:Vpush} gives the separation of the two vortices in dimensionless units; time is plotted on the bottom horizontal axis.  Fig.~\ref{fig:dpush} plots the inverse of the intervortex separation when the lower vortex unpins ($1/d_{\rm{unpin}}$) as a function of the strength of the pinning site from which it unpins.   According to Eq.~(\ref{eq:Magnusindiv}) and Eq.~(\ref{eq:magindiv}), $1/d_{\rm{unpin}}$ depends linearly on the pinning force, parametrised by $V_i$; the data in Fig.~\ref{fig:dpush} are overlaid with a linear fit (slope = 0.825) and show good agreement.

\subsection{\label{subsec:ballistic}Ballistic, unforced vortex motion}

In this subsection, we demonstrate that the vortex proximity effect is also effective when a vortex moves `naturally' along a flow-induced trajectory, ie. when it is carried ballistically by the condensate along a trajectory satisfying Schwarz's equation \citep{Barenghi}.  We contrive an experiment in which a vortex unpins (due to an increase in the global shear) and moves towards the container wall ballistically.  As the unpinned vortex approaches a second vortex, that is pinned at a larger radius, the second vortex also unpins and annihilates against the wall of the container.  

The initial conditions for this experiment are established by accelerating an $11\times11$ pinning grid from rest to $\Omega = 2.0$ instantaneously, such that 15 vortices nucleate and are pinned by the array.  We then reduce $\Omega$ to 0.075 adiabatically, causing all but six vortices to unpin and annihilate.  We focus on two vortices, both in the top right quadrant of the container, as shown in the colour plots in the top panel of Fig.~\ref{fig:proxknock}, that remain pinned within regions A and B.  The pinning in region A is then reduced to $V_i=0.4V_0$ and $\Omega$ is reduced to 0.05.  The vortex in region A unpins and moves radially outward (towards region B), causing the vortex in region B to unpin, when the intervortex separation is approximately 2.5 units.  The bottom panel of Fig.~\ref{fig:proxknock} plots $E_{\rm{latt}}$ in regions A (solid curve) and B (dashed curve) as a function of time.  The initial unpinning event in region A registers as a $\sim23\%$ increase in $E_{\rm{latt}}$ in that region; the second unpinning event in region B registers as a $\sim13\%$ increase in $E_{\rm{latt}}$ in that region.  It is unlikely that the knock-on event is caused by sound waves emitted by the travelling vortex, as we find from previous experiments (Sec.~\ref{sec:acoustic}) that the acoustic radiation from a vortex travelling with unit speed is insufficient to unpin vortices pinned with these characteristic strengths ($\sim V_0$).  We therefore conclude that the knock-on event in Fig.~\ref{fig:proxknock} is triggered by the vortex proximity effect.

The significance  of this experiment lies in two key properties.  First, by using a ballistic, uncontrived vortex trajectory (\emph{cf}. Sec.\ref{subsec:proxdrag}), we confirm that vortex-vortex repulsion is not so strong that vortices routinely \emph{self-avoid} without unpinning each other.  Second, by including additional bystander vortices (absent from Sec.~\ref{subsec:proxdrag}), we allow for the vortex lattice to stabilise the pinned state, as small adjustments in the position of pinned vortices with respect to their pinning sites can reduce the differential rotation at the site of another vortex.  Despite such stabilization, we still find that proximity knock on occurs.  This result is crucial in the many-vortex context of neutron star glitches.  

Ideally the acoustic experiments described in Sec.~\ref{sec:acoustic} would also be corroborated by repetition within a many-vortex lattice. Unfortunately, at the time of writing, this has proved impossible, because of the fine tuning of pinning strengths and global shear required.

\section{\label{sec:concindiv}Conclusions}
\begin{center}
\begin{table}[h!b!p!]
\begin{center}
\begin{tabular}{  l  l  l}
\hline\hline
  Quantity		&Value		&Units\\
\hline
  $R$ 			&$10^4$		&m			\\
  $\Omega$ 		&$10^2$		&Hz			\\
  $\rho$		&$10^{17}$	&$\rm{kg}\,\rm{m}\,\rm{s}^{-1}$ 	\\
  $\kappa$		&$10^{-8}$	&$\rm{m}^2\,\rm{s}^{-1}$		\\
  $F_{\rm{pin}}$	&$10^{12}$	&$\rm{N}\rm{m}^{-1}$  	\\
  $d_0$		&$10^{-5}$	&m  	\\
  $d_{\rm{pin}}$	&$10^{-5}$	&m 	\\
  $\Delta\Omega$	&$10^{-3}$	&Hz  	\\\hline\hline
\end{tabular}
\end{center}
\caption{Fiducial neutron star parameters.}
\label{tab:paramindiv}
\end{table}
\end{center}

The simulations presented above demonstrate that several mechanisms contribute to vortex unpinning, including the global velocity shear, sound waves (directed pulses or acoustic noise), and a vortex proximity effect due to intervortex repulsion.  The first mechanism is deterministic, and predictable if the angular acceleration of the container is known.  The second and third mechanisms potentially lead to stochastic avalanche dynamics through knock on.  The extent and timing of knock on depend unpredictably on the exact history and configuration of the vortex lattice relative to the pinning grid.  We emphasize that we observe a single generation of knock on at most; our simulated systems are presently too small to see large avalanches.  Remedying this is a crucial avenue of future work.

As discussed in Sec.~\ref{sec:intro}, a robust condensate vortex model of neutron star glitches requires a mechanism for triggering unpinning cascades if it is to explain the observed statistics and scale invariance successfully \citep{Melatos:2008p204}.   We propose that global shear is responsible for triggering such cascades, whilst the avalanche propagates via a combination of the two knock-on effects studied above.  Acoustic knock on depends on the wave damping rate (which in turn depends on the temperature).  Proximity knock on depends only on the intervortex separation (a function of $\Omega$ and the pinning strength).  

The results of this paper suggest that proximity knock on alone is insufficient to catalyse pulsar glitch avalanches, most notably because it is highly localised; the `unpinning front' quickly peters out when it hits regions of strong pinning.  In contrast, sound waves propagate throughout the system; even though the amplitude of the acoustic pulse from any given unpinning event falls away with distance from the source, it adds to the turbulent sea of acoustic noise in the system, which is perfectly capable of unpinning vortices, as we show in Sec.~\ref{subsec:sound_knock}.

Previously published models of the collective physics of neutron star glitches have successfully incorporated either one of these two effects.  \textcite{Warszawski:2008p4510} presented an avalanche model, based around nearest-neighbour interactions of the proximity type, in which long-range spatial correlations and power-law events sizes arise naturally.  The model predicts statistics consistent with astronomical data and a self-organised critical process.  The same authors \citep{Melatos:2009p4511} separately constructed a spatially homogeneous model in which each vortex experiences the same unpinning force, analogous to a homogeneous bath of sound waves.  In this coherent-noise model, pinning centres have random strengths and avalanches occur because the strength distribution of \emph{occupied} pinning sites is excavated at its lower end over time in a stochastic, history-dependent way \citep{Newman:1996p1484}.

Finally, it is instructive to make a rough quantitative comparison of the strength of acoustic and proximity knock on in the neutron star context.  Consider a neutron star with the fiducial parameters listed in Table~\ref{tab:paramindiv}, in which a single vortex unpins and travels a distance $d_0$ (the typical spacing between vortices), passing within $d_0/2$ of a nearby pinned vortex.  The additional Magnus force due to proximity of the moving vortex to the pinned vortex is $\Delta F_{\rm{M}}=\rho\kappa^2/(2\pi d_0)$.  We can convert this to an effective change in lattice energy per unit length using $\xi^2\approx10^{-12}\rm{m}^2$, obtaining $\Delta E_{\rm{latt}}\approx 10^{-7}\rm{J}\,\rm{m}^{-1}$.

By treating a condensate as a (2+1)-dimensional electrodynamic system, the acoustic power carried by sound waves is equivalent to an effective Poynting flux \citep{Lundh:2000p10757}.  GPE simulations of a vortex precessing around a pinning site at a distance $b$ give \citep{Vinen:2001p134520} 
\begin{equation}
\label{eq:power}
 P=\frac{\rho\kappa^2b^2\omega^3_{\rm{v}}}{8c_{\rm{s}}^2}~,
\end{equation}
where $\omega_{\rm{v}}$ is the precession frequency.  Qualitative evidence for acoustic radiation is found in the centre and right contour plots in Fig.~\ref{fig:diff_latt}, as well as all figures in Sec.~\ref{sec:acoustic}.  Previous studies\cite{Parker:2004p7936} found good agreement between Eq.~(\ref{eq:power}) and numerical simulations of a vortex precessing around a central impurity by comparing $E_{\rm{kin}}$ before and after sound waves are erased from a snapshot (by evolving in imaginary time).  Using Eq.~(\ref{eq:power}), and assuming that a vortex is travelling at speed $R\Delta\Omega$, the total energy per unit length emitted by the moving vortex as it traverses a distance $d_0$ is 
\begin{eqnarray}
 \Delta E_{\rm{sound}} &=& \frac{\rho \kappa^2 R^2\Omega^3}{8c_{\rm{s}}^2}\frac{d_0}{R \Delta\Omega}~.
\end{eqnarray}
For the fiducial parameters in Table~\ref{tab:paramindiv}, we find $\Delta E_{\rm{sound}}\approx 10^{-12}\rm{J}\,\rm{m}^{-1}$.  This estimate is based on a vortex moving between pinning sites, rather than during the faster, basketball-like repinning which we know produces stronger sound pulses (see Sec.~\ref{subsec:vortex_sound} and Sec.~\ref{subsec:sound_knock}).  These back-of-the-envelope calculations support our findings that the vortex proximity effect is considerably stronger than the acoustic radiation from a single moving vortex.  But of course there are many moving vortices radiating simultaneously in a large system, and their combined emission adds to the general level of acoustic noise.  A careful calculation of the radiation damping time and hence the steady-state noise level is needed to resolve the issue of which knock-on process dominates.

In conclusion, we confirm the viability of three independent mechanisms by which vortices unpin.  GPE simulations establish a qualitative microscopic basis on which a robust quantum mechanical theory of the collective physics of neutron star glitches may be constructed.\newline

\appendix

\begin{figure*}
\begin{center}
\includegraphics[scale=0.8,angle=0]{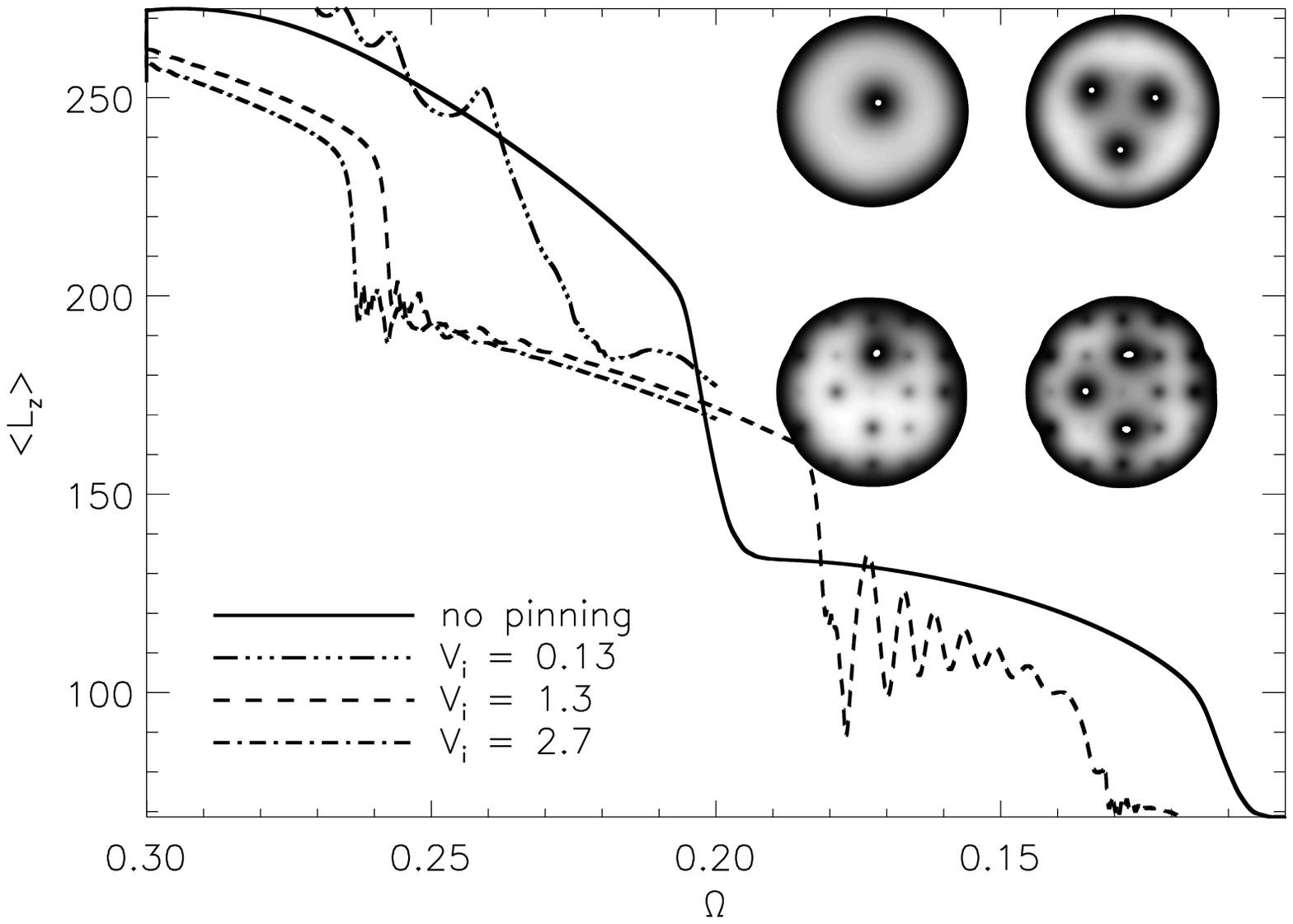}
\end{center}
\caption{Total condensate angular momentum $\langle \hat{L}_z\rangle$ as a function of container angular velocity $\Omega$ in the presence of zero ($V_i=0$, solid curve), weak ($V_i=0.13$, triple-dot-dashed curve), intermediate ($V_i=1.3$, dashed curve) and strong ($V_i=2.7$, dot-dashed curve) pinning.  \emph{Insets}:  Greyscale plots of condensate density at $t=90$ for $V_i=0$ to $V_i=2.7$ (left to right, top to bottom). The colour table runs from dark (low density) to light (high density).  Simulation parameters:  $R=8.5$, $V_{\rm{max}}=200$, $N_{\rm{c}}=10^{-3}I_{\rm{c}}$.} 
\label{fig:ch5:L_4vort}
\end{figure*}

\begin{figure}
\includegraphics[scale=0.4,angle=90]{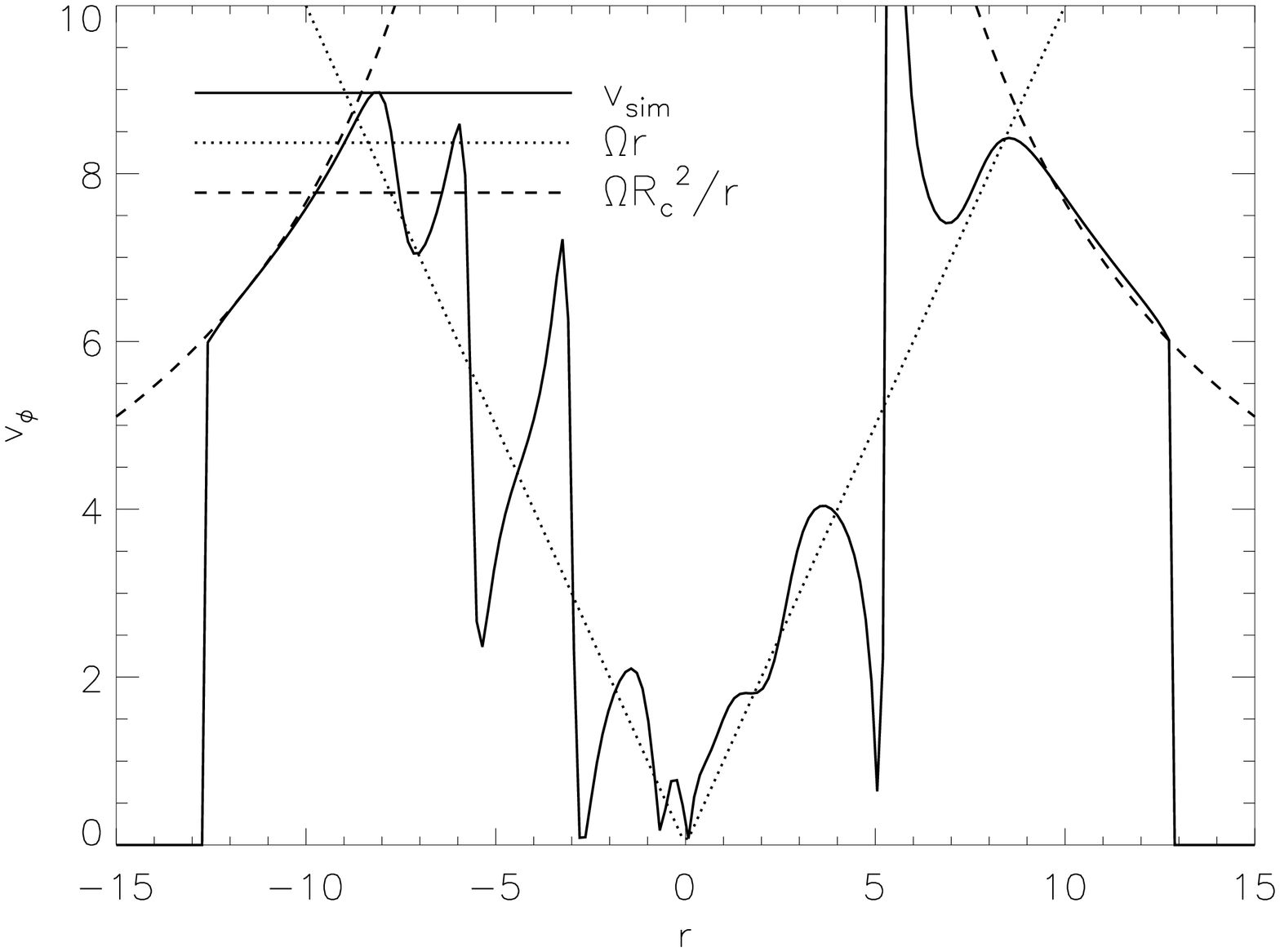}\hspace{1cm}
\includegraphics[scale=0.6,angle=90]{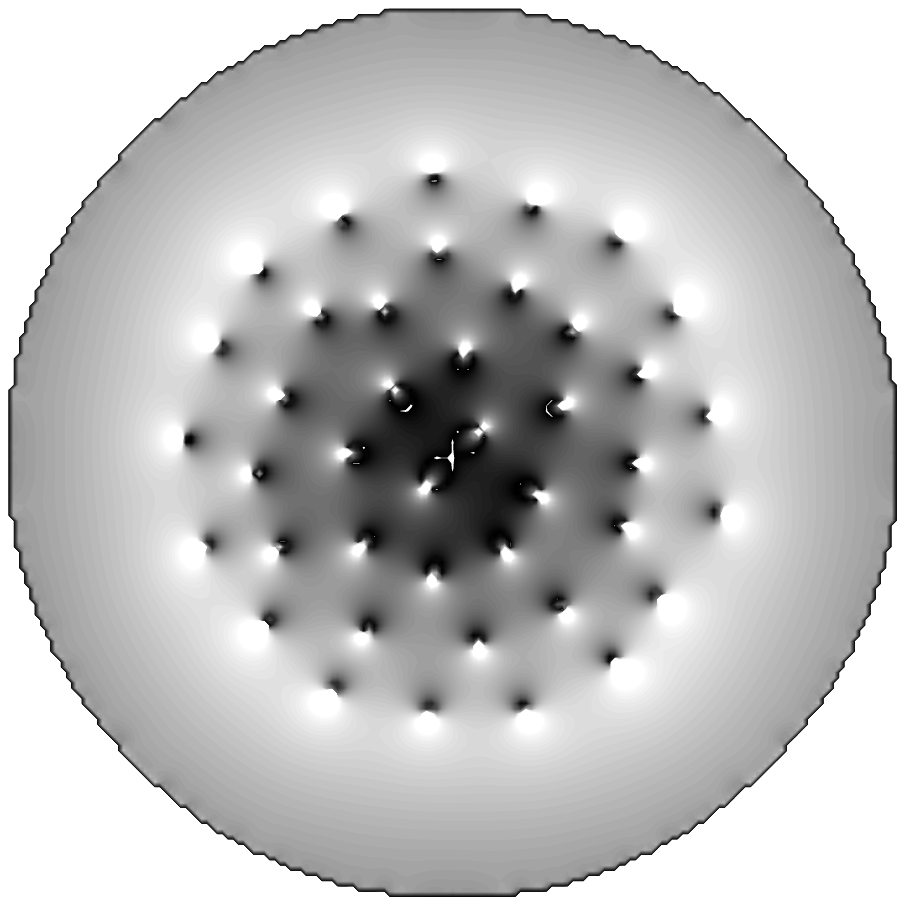}
\caption{Formation of a history-dependent core-corona structure in a vortex lattice that does not fill the condensate.  \emph{Top}:  Azimuthal velocity $v_{\phi}$ from simulation output (solid curve), rigid-body model ($v_{\phi}=\Omega r$, $\Omega=1.0$) (dotted curve), and `giant' vortex model ($v_{\phi}=\Omega R_{\rm{c}}^2/r$) (dashed curve). \emph{Bottom}:  Greyscale plot of azimuthal velocity (dark to light indicates low to high velocity); it increases smoothly with distance from the centre, punctuated by vortices.  Simulation parameters: $R=12.5$, $V_i=0.0$, $V_{\rm{max}}=200$.}
\label{fig:ch5:v}
\end{figure}

\begin{figure*}
\includegraphics[scale=0.8]{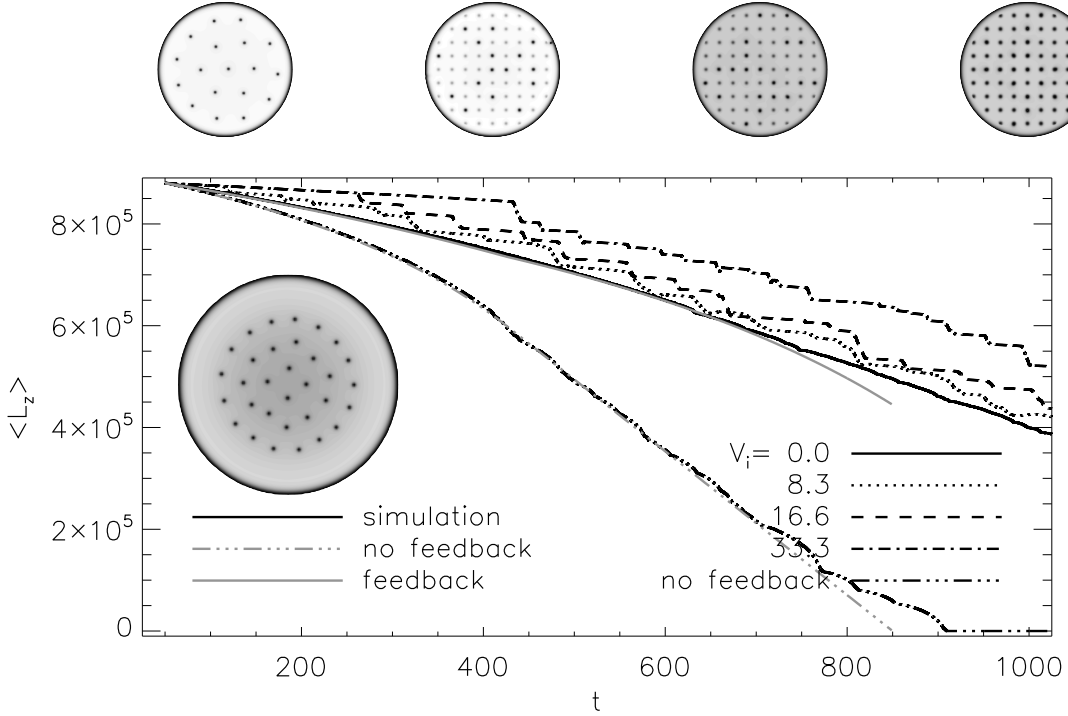}
\caption{Angular momentum $\langle \hat{L}_z\rangle$ as a function of time $t$ for spin-down experiments with pinning strengths $V_i=0.0$, 8.3, 16.6 and 33.3 (solid, dotted, dashed and dot-dashed curves respectively).  Also shown is the $V_i=0.0$ case with no feedback (triple-dot-dashed).  Overplotted are analytic predictions for the $V_i=0$ case with (solid grey) and without (triple-dot-dashed grey) feedback.  The curves are rescaled vertically to intersect at $t=0$ to correct for the fact that the initial $\langle \hat{L}_z\rangle$ varies with $V_i$ even when $\Omega$ is held constant, due to pinning hysteresis.  The greyscale plots (top) show the condensate density at $t=1000$ ($V_i=0.0$ to $V_i=33.3$, left to right).  \emph{Inset}:  the initial state for $V_i=0$, showing the vortex-free corona described in the text (see \S~\ref{subsec:corecorona}).  Simulation parameters:  $R=12.5$, $V_{\rm{max}}=200$.}
\label{fig:ch5:Lcomp}
\end{figure*} 

\begin{figure*}
\includegraphics[scale=0.8]{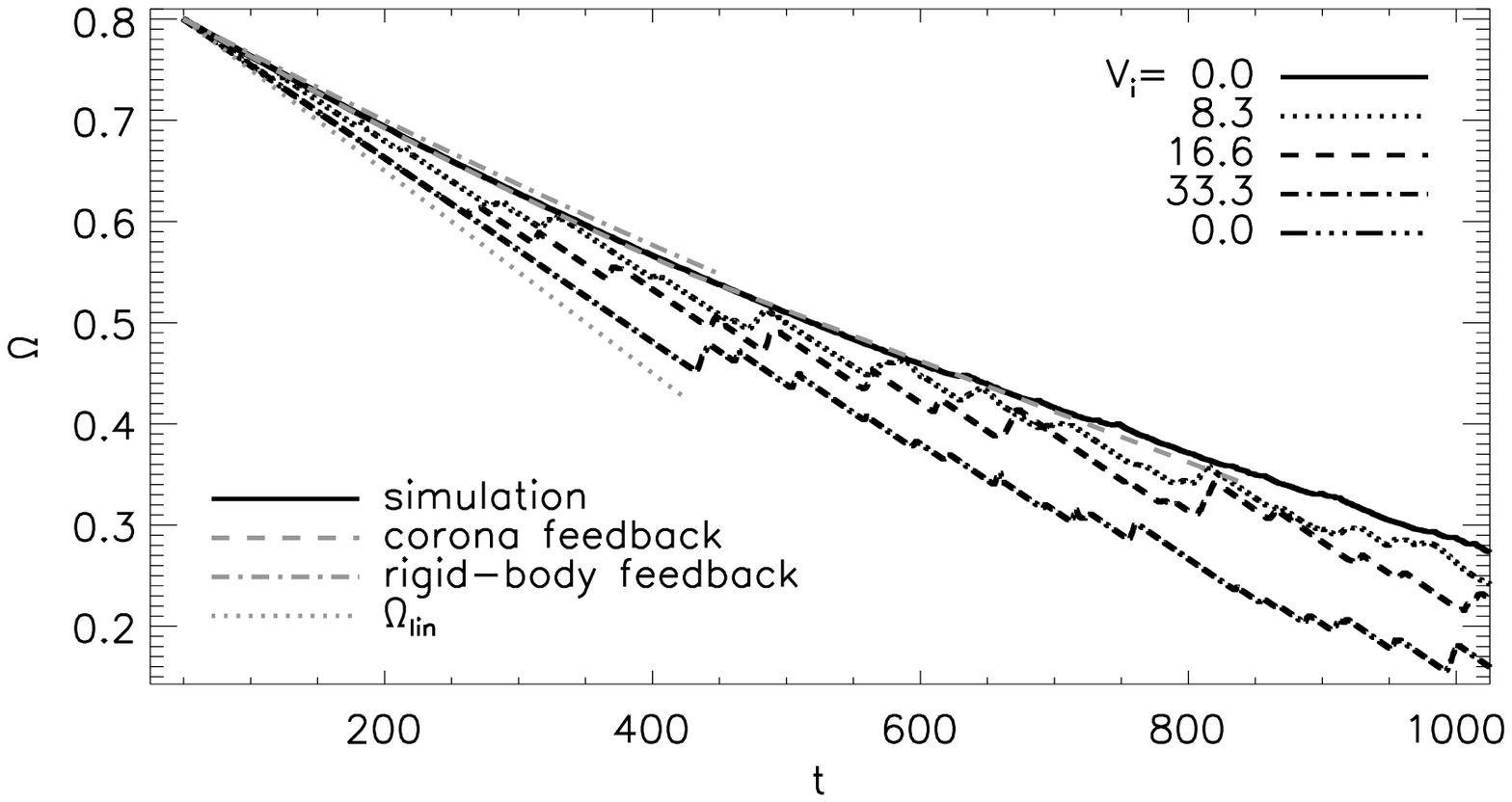}
\caption{Angular velocity $\Omega(t)$ as a function of time $t$ for spin-down experiments with pinning strengths $V_i=0.0$, 8.3, 16.6 and 33.3 (solid, dotted, dashed and dot-dashed curves respectively), corresponding to the angular momentum curves in Fig.~\ref{fig:ch5:Lcomp}.  Overplotted are analytic predictions for the $V_i=0.0$ case with feedback resulting from a condensate rotating as a rigid body (grey dot-dashed curve), feedback from a fluid with a vortex-free corona (grey dashed curve), and without feedback (dotted curve).  Simulation parameters:  $R=12.5$, $V_{\rm{max}}=200$.}
\label{fig:ch5:omcomp}
\end{figure*} 

\section{}
In this appendix, we summarize the vortex unpinning dynamics observed in 
large Gross-Pitaevskii simulations, in which multiple vortices interact 
with a large-scale grid of pinning sites, and a feedback torque acts on 
the container. The knock-on processes studied in Sec.~\ref{sec:acoustic} and \ref{sec:prox} involving
individual vortices occur repeatedly in a vortex array, but it is hard 
to isolate individual events, as their local environment is more complex.
Consequently, we track the angular-momentum-conserving response of the 
container to changes in the vortex distribution. Both the laboratory 
experiments cited in Sec.~\ref{sec:intro} and the neutron star problem involve 
$\gtrsim 10^4$ vortices.

Numerical simulations by \textcite{Sato:2007p8103}, \textcite{Yasunaga:2007p8108} and \textcite{Goldbaum:2009p8052} studied how a vortex lattice is distorted by a pinning grid (e.g. an optical lattice) for \emph{fixed} $\Omega$.  The authors Fourier transformed the condensate density to disentangle and compare the geometry of the vortex lattice and pinning grid.  They found that the transition from an Abrikosov lattice (the equilibrium configuration \citep{Chevy:2000p1060}) to a pinning-grid-like lattice is accompanied by a sharp decrease in the potential energy of the condensate.  The stronger the pinning, the more closely the vortex lattice resembles the pinning grid, and hence the lower the potential energy.  

In this appendix, we build on previous work by including the \emph{self-consistent acceleration} of the container in response to external and internal torques, to study the distortion of the vortex lattice by pinning as a function of $\Omega$.   We do so by enforcing a simple conservation rule, which assumes that any change in $\langle \hat{L}_z\rangle$ is communicated instantaneously to the container.  This approximation is valid, for example, if the torque is communicated by Kelvin waves, whose system crossing time is much shorter than the spin-down time-scale.  We also impose a constant, external spin-down torque $N_{\rm{c}}$ (electromagnetic in a pulsar, friction between the container and its supporting spindle in helium II experiments).   Hence the angular velocity $\Omega$, evolves according to the equation
\begin{eqnarray}
 I_{\rm{c}}\frac{d\Omega}{dt} &=& -\frac{d\langle \hat{L}_z\rangle}{dt}-N_{\rm{c}}~,
\label{eq:ch5:feedback}
\end{eqnarray}
where $I_{\rm c}$ is the moment of inertia of the container and $\langle \hat{L}_z\rangle$ is the expectation value of the condensate  angular momentum in the direction of the rotation axis.  $\Omega$ is the same angular velocity that appears in the GPE.  We emphasise that the observed \textit{non-glitch} spin-down rate $d\Omega/dt$ is the sum of spin down due to $N_{\rm{c}}$ and gradual decreases in  $\langle \hat{L}_z\rangle$ caused by vortices migrating to the outer edge of their pinning sites without unpinning.   The condensate does not rotate rigidly in general (especially when the number of vortices is small and/or the vortex lattice is significantly distorted by pinning), so we cannot attribute a unique angular velocity to it.

\subsection{\label{subsec:spindown_smallNv}Spasmodic spin down for small vortex number}

\begin{figure}
\begin{center}
\includegraphics[scale=.55,angle=0]{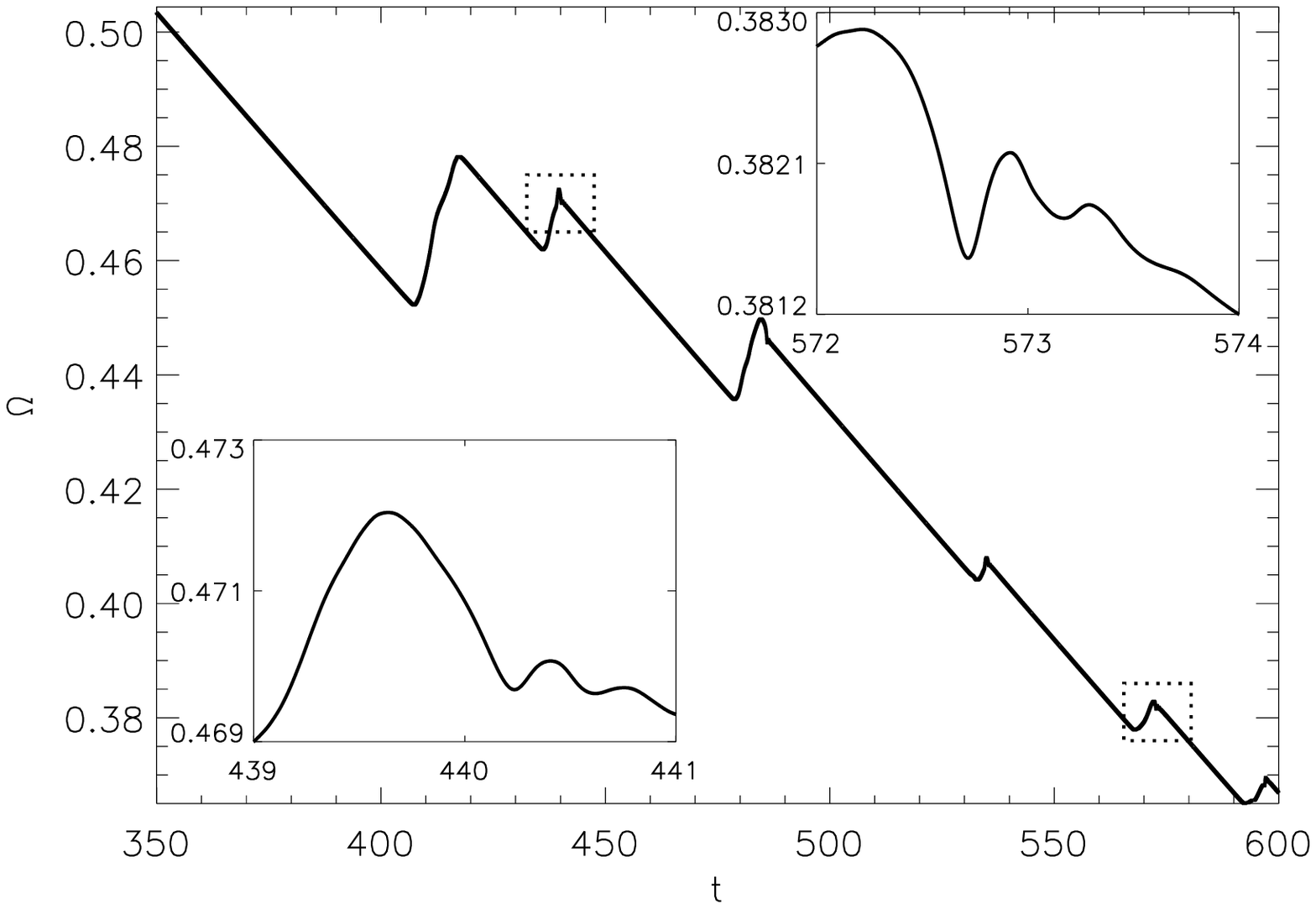}
\end{center}
\caption{Angular velocity of the container $\Omega (t)$ as a function of time $t$ for the $V_i=16.6$ experiment described in Fig.~\ref{fig:ch5:omcomp}.  The insets zoom in on the local maxima at $t=439.5$ (bottom left) and $t=572.5$ (top right), revealing oscillations following vortex repinning.  Detailed studies (see Sec.~\ref{sec:acoustic}) associate the oscillations with acoustic radiation.  Simulation parameters:  $R=12.5$, $V_{\rm{max}}=200$.}
\label{fig:ch5:close}
\end{figure}

We begin with a series of numerical experiments designed to investigate how pinning leads to `jerky' spin down of the coupled container-condensate system.  We deal first with systems containing few vortices.  As a control experiment, we impose an external torque ($N_{c}=10^{-3}I_{\rm{c}}$) on a container in which there are no pinning sites.  We set up the experiment by finding the ground state in a rotating reference frame with a $3\times3$ pinning grid; for $\Omega=0.3$ we obtain $N_{\rm{v}}=4$ vortices.  Then we remove the pinning and evolve the system to a new equilibrium, which also contains four vortices.  Initially, we switch off feedback, so that the right-hand side of Eq.~(\ref{eq:ch5:feedback}) reduces to the external torque only.  Then $\langle \hat{L}_z\rangle$ (plotted as the solid curve in Fig.~\ref{fig:ch5:L_4vort}) decreases step-wise as a function of time.  Four vortices are too few to form a uniform Abrikosov lattice, which expands homologously as the condensate decelerates.  Instead, the lattice geometry, and hence $\langle \hat{L}_z\rangle$, change abruptly as each vortex is lost.  For example, when the first vortex is lost, the lattice transforms from a square to a triangle.

Interestingly, rather than delaying deceleration by preventing vortices from moving outward, pinning appears to quicken the response of the condensate to the decelerating container for small $N_{\rm{v}}$.  That is, the condensate `feels' the deceleration of the container, because the pinning sites drag through it.  The stronger the pinning, the greater volume of condensate is displaced by the sites as they rotate with respect to the condensate, and hence the condensate responds more quickly.  The triple-dot-dashed, dot-dashed and dashed curves in Fig.~\ref{fig:ch5:L_4vort} plot $\langle \hat{L}_z\rangle(t)$ for weak, intermediate and strong pinning respectively ($V_i=0.13$, $1.3$ and $2.7$ respectively).  The loss of a vortex from the container is indicated by a step-like decrease in $\langle \hat{L}_z\rangle$.  The figure shows that the first vortex is lost earlier for larger $V_i$, demonstrating the counter-intuitive correlation described above.

In addition to discrete downward steps in $\langle \hat{L}_z\rangle$ in Fig.~\ref{fig:ch5:L_4vort}, there is also a continuous decrease prior to the first unpinning [$ \Delta \langle \hat{L}_z\rangle/\langle \hat{L}_z\rangle\approx -0.077$ from $\Omega = 0.30$ to $\Omega = 0.24$ for $V_i=1.3$].  This occurs because pinned vortices migrate towards the outer edge of their pinning sites as the global shear grows.  If we take the diameter of a pinning site to be $0.25$ (cf. $R=8.5$), positioned 2.83 units from the rotation axis (as in the top right inset of Fig.~\ref{fig:ch5:L_4vort}), the change in $\langle \hat{L}_z\rangle$ as a single vortex moves from the $b$ to $b^{\prime}$ is
\begin{eqnarray}
 \Delta \langle \hat{L}_z\rangle/\langle \hat{L}_z\rangle &=& N(b^2-b^{\prime2})/(R^2-b^2)~,
\end{eqnarray}
which gives $-0.02$ for $b=2.83$ and $b^{\prime}=3.08$.  The net change for all four original vortices is additive, i.e. $\Delta\langle \hat{L}_z\rangle/\langle \hat{L}_z\rangle\approx -0.08$, in agreement with the simulation result.

\subsection{\label{subsec:spindown_largeNv}Spasmodic spin down for large $N_{\rm{v}}$}
We now investigate the coupling of the condensate to its container as a function of pinning strength for $N_{\rm{v}}\gg 1$. The many-vortex regime is most relevant to both the laboratory and astrophysical systems discussed in \S~\ref{sec:intro}.  

\subsubsection{\label{subsec:corecorona}Core-corona structure}

According to the Feynman relation (number of vortices, $N_{\rm{v}}=2\pi R^2\Omega/\kappa$), in the absence of pinning, the vortex lattice is evenly distributed throughout the container and causes the condensate to rotate rigidly.  Figure~\ref{fig:ch5:v} demonstrates that this does not occur in general.  In plots of the azimuthal condensate velocity, $v_{\phi}=-ic_s(\psi\mathbf{\nabla}\psi^*-\psi^*\mathbf{\nabla}\psi)_{\phi}/|\psi|^2$ [cross-section and greyscale plots (the dark to light colour scale represents low to high velocity) in the \emph{top} and \emph{bottom} panels respectively of Fig.~\ref{fig:ch5:v}], two other factors modify the flow away from rigid rotation (dotted curve) in addition to a non-infinite vortex lattice.  Firstly, near vortex cores, the velocity is dominated by the $1/r$ field generated by each vortex.  Secondly, as can be seen from the lower panel of Fig.~\ref{fig:ch5:v} (see also inset in Fig.~\ref{fig:ch5:Lcomp}), the vortex lattice does not fill the trap.  This is a legacy of initialisation.  As vortex nucleation is catalysed by supercritical condensate flow past nonaxisymmetries in the container, the number of vortices nucleated in the presence of a pinning array undershoots the Feynman prediction and is hysteretic, i.e. history dependent.  In results not shown here, we find that the number of nucleated vortices for $\Omega = 1.35$ is typically 50\% smaller than the Feynman prediction.

In addition, the pinning grid perturbs the vortex lattice away from its equilibrium configuration; the condensate velocity at the trap boundary is different to the value it would take with the same number of vortices in an Abrikosov lattice.  Therefore, in order for the condensate and container rotation to match at the wall, the vortices must cluster towards the centre of the container.  In the lower panel of Fig.~\ref{fig:ch5:v}, vortices are clearly identifiable as singularities in the velocity field.  The vortex array occupies a central region $0<r<R_{\rm{c}}$ , leaving an unpopulated corona in the range $R_{\rm{c}}<r<R$.  For this vortex configuration, the condensate angular momentum, $\langle \hat{L}_z\rangle$, can be written as the sum of the contribution from the central, vortex-filled region [where the condensate rotates approximately as a rigid body, with $v_{\phi}=\Omega r$ (dotted curve)], and the vortex-free corona [where the condensate velocity field is the same as outside a single `giant' vortex, with $v_{\phi} = \Omega R_{\rm{c}}^2/r$ (dashed curve)].
Assuming that the vortices in the central region are evenly spaced, Feynman's rule gives
\begin{equation}
 R_{\rm c}^2=\frac{\kappa N_{\rm v}}{2\pi\Omega}~,
\end{equation}
leading to\footnote{We temporarily reinstate dimensions here.}
\begin{equation}
 \label{eq:ch5:Ltot}
 \langle \hat{L}_z\rangle=PQ\left(1-\frac{Q}{2\Omega R^2}\right)\,,
\end{equation}
with $P=Nm$, where $N$ is the total number of particles, and $Q=\kappa N_{\rm v}/(2\pi)$, where $\rho$ is the mean mass density (assumed uniform in these formulae).

Solving Eq.~(\ref{eq:ch5:Ltot}) and Eq.~(\ref{eq:ch5:feedback}), with $\Omega_{t=0}=\Omega_0$, 
we derive the following expression for the container angular velocity while a vortex-free corona is present and feedback is included,
\begin{eqnarray}
\label{eq:ch5:om_fb}
 \Omega(t)&=&\frac{1}{2} \Omega_{\rm{lin}}(t) -\frac{PQ^2} { 4I_{\rm{c}}R^2 \Omega_0}+\nonumber\\
	& &\left\lbrace\left[\frac{1}{2} \Omega_{\rm{lin}}(t) -\frac{PQ^2}{4I_{\rm{c}}R^2\Omega_0} \right]^2+\frac{PQ^2} { 2I_{\rm{c}}R^2 } \right\rbrace^{1/2}~,
\end{eqnarray}
where we define $\Omega_{\rm{lin}}(t)=\Omega_0-tN_{\rm{c}}/I_{\rm{c}}$.  Once the vortex lattice fills the container, condensate spin-down is assumed to mimic that of a rigid body.  It should be noted that the approximation leading to Eq.~(\ref{eq:ch5:om_fb}) assumes that $|\psi|^2$ is uniform throughout the trap.  We compare this result to simulation output in what follows.

\subsubsection{\label{subsec:largeNv}Angular momentum transport without pinning}

In this section we study $\langle \hat{L}_z\rangle(t)$ and $\Omega(t)$ for a range of $V_i$.  
The experiment is conducted on a square $200\times200$  simulation grid ($|\psi|^2=0$ outside the trap, so the simulation appears circular), with $9\times 9$ pinning sites (all $V_i$ equal, $R=12.5$).  Each simulation is run for $1000$ time units using approximately $36500$ particles.  The initial condition for each simulation is a steady-state solution for a $9\times 9$ pinning grid with $V_i=16.6$.
By way of comparison, we also present a simulation in which feedback is ignored (i.e. where changes in $\langle \hat{L}_z\rangle$ are not communicated to the container).  The results presented here are an important reminder that pinning is essential in explaining jerky condensate dynamics. 

In Fig.~\ref{fig:ch5:Lcomp} we plot $\langle \hat{L}_z\rangle(t)$ for $N_{\rm{c}}=10^{-3}I_{\rm{c}}$.  The solid black curve describes zero pinning.  As expected, $\langle \hat{L}_z\rangle$ decreases smoothly, as the vortices spread out unhindered. The triple-dot-dashed black curve repeats the zero pinning experiment but without feedback.  We observe that the condensate spins down faster without feedback, responding only to $N_{\rm{c}}$ without the spin-up torque from the decelerating condensate.  

The simulation output agrees well with the analytic prediction in the absence of feedback [Eq.~(\ref{eq:ch5:Ltot})] (triple-dot-dashed grey curve in Fig.~\ref{fig:ch5:Lcomp}) after shifting the curves to agree at $t=0$ (to correct for hysteresis in the initial vortex configuration).  When feedback is added, including the vortex-free corona, agreement between the simulation and analytic prediction (solid grey curve) remains excellent (better than one part in $10^6$) for $t<750$, at which time Eq.~(\ref{eq:ch5:om_fb}) begins to overestimate $\langle \hat{L}_z\rangle$.  This overestimate can be understood by noting that, when a vortex is about to annihilate against the wall, it accelerates radially in the region where $dV/dr$ is large and $\langle\hat{L}_z\rangle$ drops off correspondingly quickly.  Both with and without feedback, the analytic $\langle \hat{L}_z\rangle$ curves track two stages of vortex motion:  an initial stage when the vortex-free corona is present, followed by a stage during which the vortices fill the container.

\subsubsection{Angular momentum transport with pinning}

The strength of pinning determines the rate at which the condensate as a whole can decelerate: the stronger the pinning, the larger the differential rotation necessary to unpin vortices.  In the presence of a pinning lattice ($V_i>0$), the angular momentum as a function of time shows discrete downward steps.  The top row of images in Fig.~\ref{fig:ch5:Lcomp} shows the final condensate density for pinning grids with $9\times 9$ pinning sites and $V_i=0$, 8.3, 16.6 and 33.3 (solid, dotted, dashed and dot-dashed respectively).  The smaller low-density (dark) spots are unoccupied pinning sites, whereas the larger dark spots are vortices, of which there are $N_{\rm{v}}=18$, 17, 20 and 19 visible in the four images (left to right respectively).  It should be noted that, although $N_{\rm{v}}$ for $V_i=16.6$ exceeds $N_{\rm{v}}$ for $V_i=33.3$, the total angular momentum in the latter case is greater, because the vortices lie closer to the rotation axis, because they adhere more strongly to the pinning grid. 

The vortices are observed to move radially outward as the container spins down.  The motion is disjointed; \emph{vortices hop between pinning sites}.  The downward steps in angular momentum occur later for stronger pinning, as a larger differential velocity and hence Magnus force are needed to unpin.  

The step size in $\langle\hat{L}_z\rangle$ depends both on the distance travelled by the vortex before repinning (or annihilating at the wall), and the position (relative to the rotation axis) from which the vortex unpins.  For this reason we do not expect glitches of equal size.  A full explanation of how power-law distributions of glitch sizes arise in neutron stars must await the inclusion of knock-on effects.  This point is discussed in \cite{Warszawski:2010pulsar}  In summary, the effect of pinning is to create a sticky landscape for the vortices to navigate, causing the condensate to decelerate spasmodically.

\subsubsection{\label{subsubsec:response}Container response}
In Fig.~\ref{fig:ch5:omcomp}, we plot $\Omega(t)$ from simulations for the cases involving feedback in Fig.~\ref{fig:ch5:Lcomp}.  Overplotted are analytic predictions for $V_i=0$, with feedback based on a rigidly rotating condensate (grey dot dashed) and accounting for the vortex-free corona [Eq.~(\ref{eq:ch5:om_fb})] (dashed), as well as without feedback at all (dotted). The jumps in $\Omega$ accompany downward steps in $\langle \hat{L}_z\rangle$ according to Eq.~(\ref{eq:ch5:feedback}).
 
The timing of the first discrete step up in $\Omega$ scales with pinning strength.  For example, the first step up in $\Omega$ for $V_i=33.3$ occurs at $t\approx 430$, whereas for $V_i=8.3$ it occurs earlier at $t\approx120$.  

A less obvious result from these simulations is the decrease in the spin-down rate at early times (relative to spin down dictated solely by $N_{\rm{c}}$), before vortices unpin ($t\lesssim 150$).  As in the few-vortex case (Fig.~\ref{fig:ch5:L_4vort}), pinned vortices migrate to the outer edge of the pinning sites as the container decelerates, resulting in small, smooth decreases in $\langle \hat{L}_z\rangle$.  Figure~\ref{fig:ch5:omcomp} demonstrates that, prior to the first unpinning, the migration velocity of vortices is almost independent of $V_i$; for $t\lesssim 110$, all four simulation curves coincide.

In Fig.~\ref{fig:ch5:close} we zoom in on a portion of the $\Omega(t)$ curve for $V_i=16.6$.  The insets magnify the regions enclosed by the dotted rectangles.  Of particular interest is the average rise time, $\Delta t_{\rm{rise}}\approx 5$, for the `glitches'.  Possible physical time-scales associated with $\Delta t_{\rm{rise}}$ include the sound crossing time of the container ($2R/c_{\rm{s}}\approx 24$), and the sound crossing time between pinning sites ($\approx 2.5$), which overestimate and underestimate the measured $\Delta t_{\rm{rise}}$ respectively.  More promisingly, $\Delta t_{\rm{rise}}$ may be governed by the vortex travel time between pinning sites (or from pinning site to annihilation at the wall of the container).  To test this, we estimate the radial vortex velocity to be the mismatch between the azimuthal condensate velocity and the container, $v_{\rm{v}}\approx v_{\phi}-b\Omega$ ($b$ is the radius at which the vortex sits).  If we compare $\Omega(t)$ for $V_i=16.6$ with the linear ($N_{\rm{c}}$-only) spin-down curve in Fig.~\ref{fig:ch5:omcomp} at the time of the first unpinning event ($t\approx 260$), we can infer the change in $v_{\phi}$ due to vortex migration; the condensate spins down a little bit during $0<t<260$, and the container spins down a little less than it would without vortex migration.  If the vortices had not moved at all, $\Omega(t)$ and $\Omega_{\rm{lin}}$ would coincide, but instead we find $\Omega(260)-\Omega_{\rm{lin}}(260)=0.015$, which is approximately the cumulative spin-up during migration, giving $v_{\phi}-b\Omega\approx 0.175b$.  For a vortex at $b\approx 3$, we get $v_{\rm{v}}=v_{\phi}-b\Omega\approx 0.5$.  Therefore, the time for a vortex to travel between pinning sites is the inter-site separation divided by $v_{\rm{v}}$, which gives $\approx 5$, in accord with the simulation output for $\Delta t_{\rm{rise}}$.  

\begin{acknowledgments}
LW acknowledges the hospitality of the Department of Applied Mathematics and Theoretical Physics, Cambridge University, where this work was commenced.  This research was supported by an Australian Postgraduate Award.  The authors are also grateful to Andrew Martin, Cornelis A. Van Eysden and Hayder Salman for helpful discussions on many of the topics covered in this paper.
\end{acknowledgments}

\bibliography{pinning_individual}

\begin{thebibliography}{47}
\expandafter\ifx\csname natexlab\endcsname\relax\def\natexlab#1{#1}\fi
\expandafter\ifx\csname bibnamefont\endcsname\relax
  \def\bibnamefont#1{#1}\fi
\expandafter\ifx\csname bibfnamefont\endcsname\relax
  \def\bibfnamefont#1{#1}\fi
\expandafter\ifx\csname citenamefont\endcsname\relax
  \def\citenamefont#1{#1}\fi
\expandafter\ifx\csname url\endcsname\relax
  \def\url#1{\texttt{#1}}\fi
\expandafter\ifx\csname urlprefix\endcsname\relax\def\urlprefix{URL }\fi
\providecommand{\bibinfo}[2]{#2}
\providecommand{\eprint}[2][]{\url{#2}}

\bibitem[{\citenamefont{Avogadro et~al.}(2008)\citenamefont{Avogadro, Barranco,
  Broglia, and Vigezzi}}]{Avogadro:2008p29}
\bibinfo{author}{\bibfnamefont{P.}~\bibnamefont{Avogadro}},
  \bibinfo{author}{\bibfnamefont{F.}~\bibnamefont{Barranco}},
  \bibinfo{author}{\bibfnamefont{R.}~\bibnamefont{Broglia}}, \bibnamefont{and}
  \bibinfo{author}{\bibfnamefont{E.}~\bibnamefont{Vigezzi}},
  \bibinfo{journal}{Nuclear Physics A} \textbf{\bibinfo{volume}{811}},
  \bibinfo{pages}{378} (\bibinfo{year}{2008}).

\bibitem[{\citenamefont{Jones}(1997)}]{Jones:1997p26}
\bibinfo{author}{\bibfnamefont{P.~B.} \bibnamefont{Jones}},
  \bibinfo{journal}{PRL} \textbf{\bibinfo{volume}{79}}, \bibinfo{pages}{792}
  (\bibinfo{year}{1997}).

\bibitem[{\citenamefont{{Jones}}(1998)}]{Jones:1998p34}
\bibinfo{author}{\bibfnamefont{P.~B.} \bibnamefont{{Jones}}},
  \bibinfo{journal}{PRL} \textbf{\bibinfo{volume}{81}}, \bibinfo{pages}{4560}
  (\bibinfo{year}{1998}).

\bibitem[{\citenamefont{Donati and Pizzochero}(2003)}]{Donati:2003p97}
\bibinfo{author}{\bibfnamefont{P.}~\bibnamefont{Donati}} \bibnamefont{and}
  \bibinfo{author}{\bibfnamefont{P.~M.} \bibnamefont{Pizzochero}},
  \bibinfo{journal}{PRL} \textbf{\bibinfo{volume}{90}}, \bibinfo{pages}{211101}
  (\bibinfo{year}{2003}).

\bibitem[{\citenamefont{Donati and Pizzochero}(2006)}]{Donati:2006p32}
\bibinfo{author}{\bibfnamefont{P.}~\bibnamefont{Donati}} \bibnamefont{and}
  \bibinfo{author}{\bibfnamefont{P.}~\bibnamefont{Pizzochero}},
  \bibinfo{journal}{Phys. Lett. B} \textbf{\bibinfo{volume}{640}},
  \bibinfo{pages}{74} (\bibinfo{year}{2006}).

\bibitem[{\citenamefont{Avogadro et~al.}(2007)\citenamefont{Avogadro, Barranco,
  Broglia, and Vigezzi}}]{Avogadro:2007p51}
\bibinfo{author}{\bibfnamefont{P.}~\bibnamefont{Avogadro}},
  \bibinfo{author}{\bibfnamefont{F.}~\bibnamefont{Barranco}},
  \bibinfo{author}{\bibfnamefont{R.~A.} \bibnamefont{Broglia}},
  \bibnamefont{and} \bibinfo{author}{\bibfnamefont{E.}~\bibnamefont{Vigezzi}},
  \bibinfo{journal}{PRC} \textbf{\bibinfo{volume}{75}}, \bibinfo{pages}{012805}
  (\bibinfo{year}{2007}).

\bibitem[{\citenamefont{Blasio and Lazzari}(1998)}]{DeBlasio:1998p127}
\bibinfo{author}{\bibfnamefont{F.~D.} \bibnamefont{Blasio}} \bibnamefont{and}
  \bibinfo{author}{\bibfnamefont{G.}~\bibnamefont{Lazzari}},
  \bibinfo{journal}{Nuclear Physics} \textbf{\bibinfo{volume}{633}},
  \bibinfo{pages}{391} (\bibinfo{year}{1998}).

\bibitem[{\citenamefont{Jones}(2003)}]{Jones:2003p10}
\bibinfo{author}{\bibfnamefont{P.~B.} \bibnamefont{Jones}},
  \bibinfo{journal}{ApJ} \textbf{\bibinfo{volume}{595}}, \bibinfo{pages}{342}
  (\bibinfo{year}{2003}).

\bibitem[{\citenamefont{{Anderson} and {Itoh}}(1975)}]{Anderson:1975p84}
\bibinfo{author}{\bibfnamefont{P.~W.} \bibnamefont{{Anderson}}}
  \bibnamefont{and} \bibinfo{author}{\bibfnamefont{N.}~\bibnamefont{{Itoh}}},
  \bibinfo{journal}{Nature} \textbf{\bibinfo{volume}{256}}, \bibinfo{pages}{25}
  (\bibinfo{year}{1975}).

\bibitem[{\citenamefont{Alpar et~al.}(1981)\citenamefont{Alpar, Anderson,
  Pines, and Shaham}}]{Alpar:1981p18}
\bibinfo{author}{\bibfnamefont{M.~A.} \bibnamefont{Alpar}},
  \bibinfo{author}{\bibfnamefont{P.~W.} \bibnamefont{Anderson}},
  \bibinfo{author}{\bibfnamefont{D.}~\bibnamefont{Pines}}, \bibnamefont{and}
  \bibinfo{author}{\bibfnamefont{J.}~\bibnamefont{Shaham}},
  \bibinfo{journal}{PNAS} \textbf{\bibinfo{volume}{78}}, \bibinfo{pages}{5299}
  (\bibinfo{year}{1981}).

\bibitem[{\citenamefont{Dodson et~al.}(2002)\citenamefont{Dodson, McCulloch,
  and Lewis}}]{Dodson02}
\bibinfo{author}{\bibfnamefont{R.~G.} \bibnamefont{Dodson}},
  \bibinfo{author}{\bibfnamefont{P.~M.} \bibnamefont{McCulloch}},
  \bibnamefont{and} \bibinfo{author}{\bibfnamefont{D.~R.} \bibnamefont{Lewis}},
  \bibinfo{journal}{ApJ} \textbf{\bibinfo{volume}{564}}, \bibinfo{pages}{L85}
  (\bibinfo{year}{2002}).

\bibitem[{\citenamefont{Melatos et~al.}(2008)\citenamefont{Melatos, Peralta,
  and Wyithe}}]{Melatos:2008p204}
\bibinfo{author}{\bibfnamefont{A.}~\bibnamefont{Melatos}},
  \bibinfo{author}{\bibfnamefont{C.}~\bibnamefont{Peralta}}, \bibnamefont{and}
  \bibinfo{author}{\bibfnamefont{J.~S.~B.} \bibnamefont{Wyithe}},
  \bibinfo{journal}{ApJ} \textbf{\bibinfo{volume}{672}}, \bibinfo{pages}{1103}
  (\bibinfo{year}{2008}).

\bibitem[{\citenamefont{Tsakadze and Tsakadze}(1975)}]{Tsakadze:1975p5864}
\bibinfo{author}{\bibfnamefont{D.~S.} \bibnamefont{Tsakadze}} \bibnamefont{and}
  \bibinfo{author}{\bibfnamefont{S.~D.} \bibnamefont{Tsakadze}},
  \bibinfo{journal}{Journal of Experimental and Theoretical Physics Letters}
  \textbf{\bibinfo{volume}{22}}, \bibinfo{pages}{139} (\bibinfo{year}{1975}).

\bibitem[{\citenamefont{Andronikashvili
  et~al.}(1979)\citenamefont{Andronikashvili, Tsakadze, and
  Tsakadze}}]{Andronikashvili:1979p4718}
\bibinfo{author}{\bibfnamefont{E.~L.} \bibnamefont{Andronikashvili}},
  \bibinfo{author}{\bibfnamefont{J.~S.} \bibnamefont{Tsakadze}},
  \bibnamefont{and} \bibinfo{author}{\bibfnamefont{S.~J.}
  \bibnamefont{Tsakadze}}, \bibinfo{journal}{J Low Temp Phys}
  \textbf{\bibinfo{volume}{34}}, \bibinfo{pages}{13} (\bibinfo{year}{1979}).

\bibitem[{\citenamefont{Wiesenfeld et~al.}(1989)\citenamefont{Wiesenfeld, Tang,
  and Bak}}]{Wiesenfeld:1989p44}
\bibinfo{author}{\bibfnamefont{K.}~\bibnamefont{Wiesenfeld}},
  \bibinfo{author}{\bibfnamefont{C.}~\bibnamefont{Tang}}, \bibnamefont{and}
  \bibinfo{author}{\bibfnamefont{P.}~\bibnamefont{Bak}}, \bibinfo{journal}{J.
  Stat. Phys.} \textbf{\bibinfo{volume}{54}}, \bibinfo{pages}{1441}
  (\bibinfo{year}{1989}).

\bibitem[{\citenamefont{Field et~al.}(1995)\citenamefont{Field, Witt, Nori, and
  Ling}}]{Field:1995p155}
\bibinfo{author}{\bibfnamefont{S.}~\bibnamefont{Field}},
  \bibinfo{author}{\bibfnamefont{J.}~\bibnamefont{Witt}},
  \bibinfo{author}{\bibfnamefont{F.}~\bibnamefont{Nori}}, \bibnamefont{and}
  \bibinfo{author}{\bibfnamefont{X.}~\bibnamefont{Ling}},
  \bibinfo{journal}{PRL} \textbf{\bibinfo{volume}{74}}, \bibinfo{pages}{1206}
  (\bibinfo{year}{1995}).

\bibitem[{\citenamefont{Morley and Schmidt}(1996)}]{Morley:1996p2128}
\bibinfo{author}{\bibfnamefont{P.~D.} \bibnamefont{Morley}} \bibnamefont{and}
  \bibinfo{author}{\bibfnamefont{I.}~\bibnamefont{Schmidt}},
  \bibinfo{journal}{Europhysics Letters} \textbf{\bibinfo{volume}{33}},
  \bibinfo{pages}{105} (\bibinfo{year}{1996}).

\bibitem[{\citenamefont{Olson et~al.}(1996)\citenamefont{Olson, Reichhardt,
  Groth, Field, and Nori}}]{Olson:1996p3506}
\bibinfo{author}{\bibfnamefont{C.~J.} \bibnamefont{Olson}},
  \bibinfo{author}{\bibfnamefont{C.}~\bibnamefont{Reichhardt}},
  \bibinfo{author}{\bibfnamefont{J.}~\bibnamefont{Groth}},
  \bibinfo{author}{\bibfnamefont{S.}~\bibnamefont{Field}}, \bibnamefont{and}
  \bibinfo{author}{\bibfnamefont{F.}~\bibnamefont{Nori}}, \bibinfo{journal}{APS
  Meeting Abstracts} p. \bibinfo{pages}{1103} (\bibinfo{year}{1996}).

\bibitem[{\citenamefont{{Bassler} and {Paczuski}}(1998)}]{Bassler:1998p8}
\bibinfo{author}{\bibfnamefont{K.~E.} \bibnamefont{{Bassler}}}
  \bibnamefont{and}
  \bibinfo{author}{\bibfnamefont{M.}~\bibnamefont{{Paczuski}}},
  \bibinfo{journal}{PRL} \textbf{\bibinfo{volume}{81}}, \bibinfo{pages}{3761}
  (\bibinfo{year}{1998}).

\bibitem[{\citenamefont{Warszawski and Melatos}(2008)}]{Warszawski:2008p4510}
\bibinfo{author}{\bibfnamefont{L.}~\bibnamefont{Warszawski}} \bibnamefont{and}
  \bibinfo{author}{\bibfnamefont{A.}~\bibnamefont{Melatos}},
  \bibinfo{journal}{MNRAS} \textbf{\bibinfo{volume}{390}}, \bibinfo{pages}{175}
  (\bibinfo{year}{2008}).

\bibitem[{\citenamefont{Sato et~al.}(2007)\citenamefont{Sato, Ishiyama, and
  Nikuni}}]{Sato:2007p8103}
\bibinfo{author}{\bibfnamefont{T.}~\bibnamefont{Sato}},
  \bibinfo{author}{\bibfnamefont{T.}~\bibnamefont{Ishiyama}}, \bibnamefont{and}
  \bibinfo{author}{\bibfnamefont{T.}~\bibnamefont{Nikuni}},
  \bibinfo{journal}{PRA} \textbf{\bibinfo{volume}{76}}, \bibinfo{pages}{053628}
  (\bibinfo{year}{2007}).

\bibitem[{\citenamefont{Yasunaga and Tsubota}(2007)}]{Yasunaga:2007p8108}
\bibinfo{author}{\bibfnamefont{M.}~\bibnamefont{Yasunaga}} \bibnamefont{and}
  \bibinfo{author}{\bibfnamefont{M.}~\bibnamefont{Tsubota}},
  \bibinfo{journal}{J Low Temp Phys} \textbf{\bibinfo{volume}{148}},
  \bibinfo{pages}{363} (\bibinfo{year}{2007}).

\bibitem[{\citenamefont{Hakonen et~al.}(1998)\citenamefont{Hakonen, Avenel, and
  Varoquaux}}]{Hakonen:1998p10787}
\bibinfo{author}{\bibfnamefont{P.}~\bibnamefont{Hakonen}},
  \bibinfo{author}{\bibfnamefont{O.}~\bibnamefont{Avenel}}, \bibnamefont{and}
  \bibinfo{author}{\bibfnamefont{E.}~\bibnamefont{Varoquaux}},
  \bibinfo{journal}{PRL} \textbf{\bibinfo{volume}{81}}, \bibinfo{pages}{3451}
  (\bibinfo{year}{1998}).

\bibitem[{\citenamefont{Varoquaux et~al.}(1998)\citenamefont{Varoquaux, Avenel,
  and Hakonen}}]{Varoquaux:1998p2402}
\bibinfo{author}{\bibfnamefont{E.}~\bibnamefont{Varoquaux}},
  \bibinfo{author}{\bibfnamefont{O.}~\bibnamefont{Avenel}}, \bibnamefont{and}
  \bibinfo{author}{\bibfnamefont{P.}~\bibnamefont{Hakonen}},
  \bibinfo{journal}{Physica B: Physics of Condensed Matter}
  \textbf{\bibinfo{volume}{255}} (\bibinfo{year}{1998}).

\bibitem[{\citenamefont{Varoquaux et~al.}(2000)\citenamefont{Varoquaux, Avenel,
  Hakonen, and Mukharsky}}]{Varoquaux200087}
\bibinfo{author}{\bibfnamefont{E.}~\bibnamefont{Varoquaux}},
  \bibinfo{author}{\bibfnamefont{O.}~\bibnamefont{Avenel}},
  \bibinfo{author}{\bibfnamefont{P.}~\bibnamefont{Hakonen}}, \bibnamefont{and}
  \bibinfo{author}{\bibfnamefont{Y.}~\bibnamefont{Mukharsky}},
  \bibinfo{journal}{Physica B: Physics of Condensed Matter}
  \textbf{\bibinfo{volume}{284-288}}, \bibinfo{pages}{87}
  (\bibinfo{year}{2000}).

\bibitem[{\citenamefont{Link}(2009)}]{Link:2009p9063}
\bibinfo{author}{\bibfnamefont{B.}~\bibnamefont{Link}}, \bibinfo{journal}{PRL}
  \textbf{\bibinfo{volume}{102}}, \bibinfo{pages}{131101}
  (\bibinfo{year}{2009}).

\bibitem[{\citenamefont{{Sedrakian}}(1995)}]{Sedrakian:1995MNRAS}
\bibinfo{author}{\bibfnamefont{A.~D.} \bibnamefont{{Sedrakian}}},
  \bibinfo{journal}{MNRAS} \textbf{\bibinfo{volume}{277}}, \bibinfo{pages}{225}
  (\bibinfo{year}{1995}).

\bibitem[{\citenamefont{{Hills} and {Roberts}}(1977)}]{Hills:1977}
\bibinfo{author}{\bibfnamefont{R.~N.} \bibnamefont{{Hills}}} \bibnamefont{and}
  \bibinfo{author}{\bibfnamefont{P.~H.} \bibnamefont{{Roberts}}},
  \bibinfo{journal}{Archive for Rational Mechanics and Analysis}
  \textbf{\bibinfo{volume}{66}}, \bibinfo{pages}{43} (\bibinfo{year}{1977}).

\bibitem[{\citenamefont{Penckwitt et~al.}(2002)\citenamefont{Penckwitt,
  Ballagh, and Gardiner}}]{Penckwitt:2002p1045}
\bibinfo{author}{\bibfnamefont{A.~A.} \bibnamefont{Penckwitt}},
  \bibinfo{author}{\bibfnamefont{R.~J.} \bibnamefont{Ballagh}},
  \bibnamefont{and} \bibinfo{author}{\bibfnamefont{C.~W.}
  \bibnamefont{Gardiner}}, \bibinfo{journal}{PRL}
  \textbf{\bibinfo{volume}{89}}, \bibinfo{pages}{260402}
  (\bibinfo{year}{2002}).

\bibitem[{\citenamefont{{Roberts} and {Berloff}}(2001)}]{Roberts:2001LNP}
\bibinfo{author}{\bibfnamefont{P.~H.} \bibnamefont{{Roberts}}}
  \bibnamefont{and} \bibinfo{author}{\bibfnamefont{N.~G.}
  \bibnamefont{{Berloff}}}, in \emph{\bibinfo{booktitle}{Quantized Vortex
  Dynamics and Superfluid Turbulence}}, edited by
  \bibinfo{editor}{\bibnamefont{{C.~F.~Barenghi, R.~J.~Donnelly, \&
  W.~F.~Vinen}}} (\bibinfo{year}{2001}), vol. \bibinfo{volume}{571} of
  \emph{\bibinfo{series}{Lecture Notes in Physics, Berlin Springer Verlag}},
  pp. \bibinfo{pages}{235--+}.

\bibitem[{\citenamefont{{Pomeau} and {Rica}}(1993)}]{Pomeau:1993PRL}
\bibinfo{author}{\bibfnamefont{Y.}~\bibnamefont{{Pomeau}}} \bibnamefont{and}
  \bibinfo{author}{\bibfnamefont{S.}~\bibnamefont{{Rica}}},
  \bibinfo{journal}{PRL} \textbf{\bibinfo{volume}{71}}, \bibinfo{pages}{247}
  (\bibinfo{year}{1993}).

\bibitem[{\citenamefont{Choi et~al.}(1998)\citenamefont{Choi, Morgan, and
  Burnett}}]{Choi:1998p9985}
\bibinfo{author}{\bibfnamefont{S.}~\bibnamefont{Choi}},
  \bibinfo{author}{\bibfnamefont{S.~A.} \bibnamefont{Morgan}},
  \bibnamefont{and} \bibinfo{author}{\bibfnamefont{K.}~\bibnamefont{Burnett}},
  \bibinfo{journal}{PRA} \textbf{\bibinfo{volume}{57}}, \bibinfo{pages}{4057}
  (\bibinfo{year}{1998}).

\bibitem[{\citenamefont{Tsubota et~al.}(2002)\citenamefont{Tsubota, Kasamatsu,
  and Ueda}}]{Tsubota:2002p11}
\bibinfo{author}{\bibfnamefont{M.}~\bibnamefont{Tsubota}},
  \bibinfo{author}{\bibfnamefont{K.}~\bibnamefont{Kasamatsu}},
  \bibnamefont{and} \bibinfo{author}{\bibfnamefont{M.}~\bibnamefont{Ueda}},
  \bibinfo{journal}{PRA} \textbf{\bibinfo{volume}{65}} (\bibinfo{year}{2002}).

\bibitem[{\citenamefont{{Kasamatsu} et~al.}(2003)\citenamefont{{Kasamatsu},
  {Tsubota}, and {Ueda}}}]{Kasamatsu:2003p1051}
\bibinfo{author}{\bibfnamefont{K.}~\bibnamefont{{Kasamatsu}}},
  \bibinfo{author}{\bibfnamefont{M.}~\bibnamefont{{Tsubota}}},
  \bibnamefont{and} \bibinfo{author}{\bibfnamefont{M.}~\bibnamefont{{Ueda}}},
  \bibinfo{journal}{PRA} \textbf{\bibinfo{volume}{67}}, \bibinfo{pages}{033610}
  (\bibinfo{year}{2003}).

\bibitem[{\citenamefont{{Mewes} et~al.}(1996)\citenamefont{{Mewes}, {Andrews},
  {van Druten}, {Kurn}, {Durfee}, and {Ketterle}}}]{Mewes:1996p9989}
\bibinfo{author}{\bibfnamefont{M.~O.} \bibnamefont{{Mewes}}},
  \bibinfo{author}{\bibfnamefont{M.~R.} \bibnamefont{{Andrews}}},
  \bibinfo{author}{\bibfnamefont{N.~J.} \bibnamefont{{van Druten}}},
  \bibinfo{author}{\bibfnamefont{D.~M.} \bibnamefont{{Kurn}}},
  \bibinfo{author}{\bibfnamefont{D.~S.} \bibnamefont{{Durfee}}},
  \bibnamefont{and}
  \bibinfo{author}{\bibfnamefont{W.}~\bibnamefont{{Ketterle}}},
  \bibinfo{journal}{PRL} \textbf{\bibinfo{volume}{77}}, \bibinfo{pages}{416}
  (\bibinfo{year}{1996}).

\bibitem[{\citenamefont{Jackson and Zaremba}(2001)}]{Jackson:2001p10058}
\bibinfo{author}{\bibfnamefont{B.}~\bibnamefont{Jackson}} \bibnamefont{and}
  \bibinfo{author}{\bibfnamefont{E.}~\bibnamefont{Zaremba}},
  \bibinfo{journal}{PRL} \textbf{\bibinfo{volume}{87}}, \bibinfo{pages}{100404}
  (\bibinfo{year}{2001}).

\bibitem[{\citenamefont{{Jackson} et~al.}(2009)\citenamefont{{Jackson},
  {Proukakis}, {Barenghi}, and {Zaremba}}}]{Jackson:2009p8818}
\bibinfo{author}{\bibfnamefont{B.}~\bibnamefont{{Jackson}}},
  \bibinfo{author}{\bibfnamefont{N.~P.} \bibnamefont{{Proukakis}}},
  \bibinfo{author}{\bibfnamefont{C.~F.} \bibnamefont{{Barenghi}}},
  \bibnamefont{and}
  \bibinfo{author}{\bibfnamefont{E.}~\bibnamefont{{Zaremba}}},
  \bibinfo{journal}{PRA} \textbf{\bibinfo{volume}{79}}, \bibinfo{pages}{053615}
  (\bibinfo{year}{2009}).

\bibitem[{\citenamefont{Donnelly}(1991)}]{Donnelly}
\bibinfo{author}{\bibfnamefont{R.~J.} \bibnamefont{Donnelly}},
  \emph{\bibinfo{title}{Quantized vortices in Helium II}}
  (\bibinfo{publisher}{Cambridge University Press}, \bibinfo{year}{1991}).

\bibitem[{\citenamefont{Barenghi et~al.}(2001)\citenamefont{Barenghi, Donnelly,
  and \emph{Eds}}}]{Barenghi}
\bibinfo{author}{\bibfnamefont{C.~F.} \bibnamefont{Barenghi}},
  \bibinfo{author}{\bibfnamefont{R.~J.} \bibnamefont{Donnelly}},
  \bibnamefont{and} \bibinfo{author}{\bibfnamefont{W.~F.~V.}
  \bibnamefont{\emph{Eds}}}, \emph{\bibinfo{title}{Quantized vortex dynamics
  and superfluid turbulence}} (\bibinfo{publisher}{Physics and Astronomy,
  Springer}, \bibinfo{year}{2001}).

\bibitem[{\citenamefont{Melatos and Warszawski}(2009)}]{Melatos:2009p4511}
\bibinfo{author}{\bibfnamefont{A.}~\bibnamefont{Melatos}} \bibnamefont{and}
  \bibinfo{author}{\bibfnamefont{L.}~\bibnamefont{Warszawski}},
  \bibinfo{journal}{ApJ} \textbf{\bibinfo{volume}{700}}, \bibinfo{pages}{1524}
  (\bibinfo{year}{2009}).

\bibitem[{\citenamefont{Newman and Sneppen}(1996)}]{Newman:1996p1484}
\bibinfo{author}{\bibfnamefont{M.}~\bibnamefont{Newman}} \bibnamefont{and}
  \bibinfo{author}{\bibfnamefont{K.}~\bibnamefont{Sneppen}},
  \bibinfo{journal}{PRE} \textbf{\bibinfo{volume}{54}} (\bibinfo{year}{1996}).

\bibitem[{\citenamefont{Lundh and Ao}(2000)}]{Lundh:2000p10757}
\bibinfo{author}{\bibfnamefont{E.}~\bibnamefont{Lundh}} \bibnamefont{and}
  \bibinfo{author}{\bibfnamefont{P.}~\bibnamefont{Ao}},
  \bibinfo{journal}{Physica B: Condensed Matter} \textbf{\bibinfo{volume}{284}}
  (\bibinfo{year}{2000}).

\bibitem[{\citenamefont{Vinen}(2001)}]{Vinen:2001p134520}
\bibinfo{author}{\bibfnamefont{W.~F.} \bibnamefont{Vinen}},
  \bibinfo{journal}{PRB} \textbf{\bibinfo{volume}{64}}, \bibinfo{pages}{134520}
  (\bibinfo{year}{2001}).

\bibitem[{\citenamefont{Parker et~al.}(2004)\citenamefont{Parker, Proukakis,
  Barenghi, and Adams}}]{Parker:2004p7936}
\bibinfo{author}{\bibfnamefont{N.~G.} \bibnamefont{Parker}},
  \bibinfo{author}{\bibfnamefont{N.~P.} \bibnamefont{Proukakis}},
  \bibinfo{author}{\bibfnamefont{C.~F.} \bibnamefont{Barenghi}},
  \bibnamefont{and} \bibinfo{author}{\bibfnamefont{C.~S.} \bibnamefont{Adams}},
  \bibinfo{journal}{PRL} \textbf{\bibinfo{volume}{92}}, \bibinfo{pages}{160403}
  (\bibinfo{year}{2004}).

\bibitem[{\citenamefont{Goldbaum and Mueller}(2009)}]{Goldbaum:2009p8052}
\bibinfo{author}{\bibfnamefont{D.~S.} \bibnamefont{Goldbaum}} \bibnamefont{and}
  \bibinfo{author}{\bibfnamefont{E.~J.} \bibnamefont{Mueller}},
  \bibinfo{journal}{PRA} \textbf{\bibinfo{volume}{79}}, \bibinfo{pages}{063625}
  (\bibinfo{year}{2009}).

\bibitem[{\citenamefont{Chevy et~al.}(2000)\citenamefont{Chevy, Madison, and
  Dalibard}}]{Chevy:2000p1060}
\bibinfo{author}{\bibfnamefont{F.}~\bibnamefont{Chevy}},
  \bibinfo{author}{\bibfnamefont{K.~W.} \bibnamefont{Madison}},
  \bibnamefont{and} \bibinfo{author}{\bibfnamefont{J.}~\bibnamefont{Dalibard}},
  \bibinfo{journal}{PRL} \textbf{\bibinfo{volume}{85}}, \bibinfo{pages}{2223}
  (\bibinfo{year}{2000}).

\bibitem[{\citenamefont{{Warszawski} and
  {Melatos}}(2011)}]{Warszawski:2010pulsar}
\bibinfo{author}{\bibfnamefont{L.}~\bibnamefont{{Warszawski}}}
  \bibnamefont{and}
  \bibinfo{author}{\bibfnamefont{A.}~\bibnamefont{{Melatos}}},
  \bibinfo{journal}{MNRAS} pp. \bibinfo{pages}{767--+} (\bibinfo{year}{2011}).

\end{thebibliography}

\end{document}